\documentclass[preprint2]{aastex631}

\usepackage{comment}
\usepackage{amsmath}

\begin{document}

\title{Atmospheric Response for MeV $\mathrm{\gamma}$ Rays Observed with Balloon-Borne Detectors}

\author[0000-0002-6774-3111]{Christopher M. Karwin}
\affiliation{NASA Postdoctoral Program Fellow} \email{christopher.m.karwin@nasa.gov}
\affiliation{NASA Goddard Space Flight Center, Greenbelt, MD, 20771, USA}

\author[0000-0001-6677-914X]{Carolyn Kierans}
\affiliation{NASA Goddard Space Flight Center, Greenbelt, MD, 20771, USA}

\author[0000-0001-6874-2594]{Albert Y. Shih}
\affiliation{NASA Goddard Space Flight Center, Greenbelt, MD, 20771, USA}

\author[0000-0002-2471-8696]{Israel Martinez Castellanos}
\affiliation{NASA Goddard Space Flight Center, Greenbelt, MD, 20771, USA}
\affiliation{Department of Astronomy, University of Maryland, College Park, Maryland 20742, USA}

\author{Alex Lowell}
\affiliation{Space Sciences Laboratory, UC Berkeley, 7 Gauss Way, Berkeley, CA 94720, USA}

\author[0000-0002-0552-3535]{Thomas Siegert}
\affiliation{Julius-Maximilians-Universität Würzburg, Fakultät für Physik und Astronomie, Institut für Theoretische Physik und Astrophysik, Lehrstuhl für Astronomie, Emil-Fischer-Str. 31, D-97074 Würzburg, Germany}

\author{Jarred Roberts}
\affiliation{Department of Astronomy \& Astrophysics, UC San Diego, 9500 Gilman Drive, La Jolla CA 92093, USA}

\author[0000-0002-2664-8804]{Savitri Gallego}
\affiliation{Institut für Physik \& Exzellenzcluster PRISMA+, Johannes Gutenberg-Universität Mainz, 55099 Mainz, Germany}

\author[0000-0003-1521-7950]{Adrien Laviron}
\affiliation{Laboratoire Leprince-Ringuet, CNRS/IN2P3, École polytechnique, Institut Polytechnique de Paris, 91120 Palaiseau, France}

\author[0000-0001-9067-3150]{Andreas Zoglauer}
\affiliation{Space Sciences Laboratory, UC Berkeley, 7 Gauss Way, Berkeley, CA 94720, USA}

\author[0000-0001-5506-9855]{John A. Tomsick}
\affiliation{Space Sciences Laboratory, UC Berkeley, 7 Gauss Way, Berkeley, CA 94720, USA}

\author[0000-0001-9567-4224]{Steven E. Boggs}
\affiliation{Department of Astronomy \& Astrophysics, UC San Diego, 9500 Gilman Drive, La Jolla CA 92093, USA}

\collaboration{12}{(for the COSI Collaboration)}



\begin{abstract}

The atmospheric response for MeV $\gamma$ rays ($\sim$ 0.1 $-$ 10 MeV) can be characterized in terms of two observed components. The first component is due to photons that reach the detector without scattering. The second component is due to photons that reach the detector after scattering one or more times. While the former can be determined in a straightforward manner, the latter is much more complex to quantify, as it requires tracking the transport of all source photons that are incident on Earth's atmosphere. The scattered component can cause a significant energy-dependent distortion in the measured spectrum, which is important to account for when making balloon-borne observations. In this work we simulate the full response for $\gamma$-ray transport in the atmosphere. We find that the scattered component becomes increasingly more significant towards lower energies, and at 0.1 MeV it may increase the measured flux by as much as a factor of $\sim2-4$, depending on the photon index and off-axis angle of the source. This is particularly important for diffuse sources, whereas the effect from scattering can be significantly reduced for point sources observed with an imaging telescope.   
\end{abstract}



\section{Introduction} \label{sec:intro}
As $\gamma$ rays with energies between $\sim$ 0.1 $-$ 10 MeV travel through Earth's atmosphere, they may undergo Compton scattering. This causes attenuation and distortion of the original signal. In order to overcome these atmospheric effects, observations in this energy band are typically made with space-based telescopes. However, balloon-borne observations within Earth's atmosphere are still essential for the development of new telescope technologies. Such development is quickly progressing, as exemplified by the long-duration balloon flights of the Compton Spectrometer and Imager (COSI)~\citep{Kierans:2016qik,COSIofficial,BeechertCOSICalib,Tomsick:2023aue} and the Sub-MeV/MeV gamma-ray Imaging Loaded-on-balloon Experiments (SMILE)~\citep{2022ApJ...930....6T}, as well as the recent short-duration balloon flight of the Compton Pair telescope (ComPair)~\citep{2023arXiv230812464V}, and the upcoming balloon flight of the Antarctic Demonstrator for the Advanced Particle-astrophysics Telescope (ADAPT)~\citep{Chen:2023nij}. Moreover, a number of other mission concepts have recently been proposed, and may very well undergo balloon tests in the coming years, e.g., the All-sky Medium Energy Gamma-ray Observatory eXplorer (AMEGO-X)~\citep{Caputo:2022xpx} and the Galactic Explorer with a Coded aperture mask Compton telescope (GECCO)~\citep{2022JCAP...07..036O}. In order to make an accurate assessment of a telescope's performance when operating at balloon altitudes, it is imperative to have a complete understanding of how the atmosphere affects $\gamma$-ray transport. Specifically, in this work we focus on the spectral response. 

The scattering of MeV photons in the atmosphere can be characterized in terms of two components. First, only a fraction of photons from a source will travel through the atmosphere and reach the detector without scattering, which we refer to as the transmitted photons. This component can be accounted for in a straightforward manner by calculating the corresponding transmission probability (TP). Generally, the TP depends on the initial energy of the photon, the off-axis angle of the source, and the altitude of the observations. Second, some fraction of photons from a source will reach the detector after one or more scatters (even if not initially directed towards the detector), which we refer to as the scattered photons. Accounting for this component is much more challenging, as it requires tracking the $\gamma$-ray transport in the atmosphere for all incident photons. Moreover, $\gamma$ rays from astrophysical sources pass the Earth as plane waves, and in principle this implies that the scattered photons can come from a surface area effectively as large as the cross-section of Earth's upper atmosphere, which far exceeds the area of a detector. This can lead to difficulties when it comes to acquiring appropriate statistics from Monte Carlo simulations of a detector's response.    

The scattered component is important to account for when analyzing balloon-borne observations, as it can produce a significant energy-dependent distortion in the measured spectrum. In general, this tends to lead to more photons towards lower energies. For diffuse continuum sources, such as the Galactic diffuse continuum emission and extragalactic $\gamma$-ray background, photons enter the detector from all directions, and therefore the spectral distortion from the scattered photons becomes very significant. The effect is not as crucial for point sources observed with imaging telescopes, such as COSI, because they have the ability to decipher the photon's incident direction. However, the degree to which photons can be selected coming from the source direction also has a strong dependence on the instrument's angular resolution.

There are only a small number of previous works which have made detailed estimates of the spectral response from atmospheric scattering at balloon altitudes. Of particular note is the work presented in~\citet{2011ApJ...733...13T}, in which they were able to successfully fit the
growth curve of the SMILE balloon flight. The analysis includes a direct estimate of the scattered component, which is also compared to a few older works, calculated either analytically or from simulations (see~\citet{2011ApJ...733...13T} and references therein). Another notable work is the balloon-borne measurements of the supernova SN 1987A in the hard X-ray continuum, as presented in~\citet{1995ApJ...439..963P}. The analysis includes corrections for atmospheric absorption and scattering, which were estimated from simulations, based on a rectangular mass model of Earth's atmosphere. More recently,~\citet{2021arXiv210509524P} calculated atmospheric response matrices, although the focus of that work was on the reflection component for GRBs observed with space-based observationss.

In this work we determine the full response for $\gamma$-ray transport in the atmosphere via Monte Carlo simulations, which includes both the transmitted and scattered components. As part of this, we publicly release the COSI atmosphere simulation and analysis pipeline, cosi-atmosphere\footnote{The cosi-atmosphere package is available at \url{https://cosi-atmosphere.readthedocs.io/en/latest/}}~\citep{karwin_2024_12668447}. Additionally, we provide atmospheric response matrices calculated for altitudes between 25.5 $-$ 40.5 km (in 1 km steps)\footnote{Instructions for accessing the atmospheric response matrices can be found in the cosi-atmosphere documentation.}. The goal of the cosi-atmosphere package is to build a user-friendly Python-based library that can be employed for atmospheric physics associated with MeV $\gamma$-ray astronomy. Note that these tools are independent of any specific detector, and thus they can be easily adapted for different observations. The package currently includes the spectral response for balloon-borne observations, and we plan to extend the tools to other relevant topics in the near future, including the determination of atmospheric $\gamma$-ray backgrounds, which dominant the emission at balloon-altitudes, as well as determination of the $\gamma$-ray albedo and reflection components, for space-based observations.

The paper proceeds as follows. In Section~\ref{sec:sims} we detail the simulation setup, which is based on a spherical geometry. In Section~\ref{sec:response} we describe the atmospheric response, including some specific applications, and validation of the simulations. The summary and conclusions are given in Section~\ref{sec:conclusions}. In Appendix~\ref{sec:photon_distributions} we give more details about the simulations. Details regarding the interactions that occur during the $\gamma$-ray transport through the atmosphere are given in Appendix~\ref{sec:photon_interactions}. In Appendix~\ref{sec:tp_analytical} we give details for calculating the TP analytically. Appendix~\ref{sec:on_axis_spherical} provides an additional example complementing Section~\ref{sec:response}. In Appendix~\ref{sec:rectangular_mass_model} we present response calculations using a simplified model of the atmosphere, consisting of a rectangular geometry, which we show to be consistent with the spherical geometry simulations. In order to use a concrete example, our calculations in this work are based on the 2016 COSI balloon flight~\citep{Kierans:2016qik}, which primarily motivates our choice of inputs for the atmospheric model. Specifically, we use a representative date and geographical location of 2016-06-13 and $(\mathrm{lat,lon})=(-5.66^\circ,-107.38^\circ)$, respectively.   

\section{Simulations} \label{sec:sims}

\begin{figure*}[t]
\centering
\includegraphics[width=2.0\columnwidth]{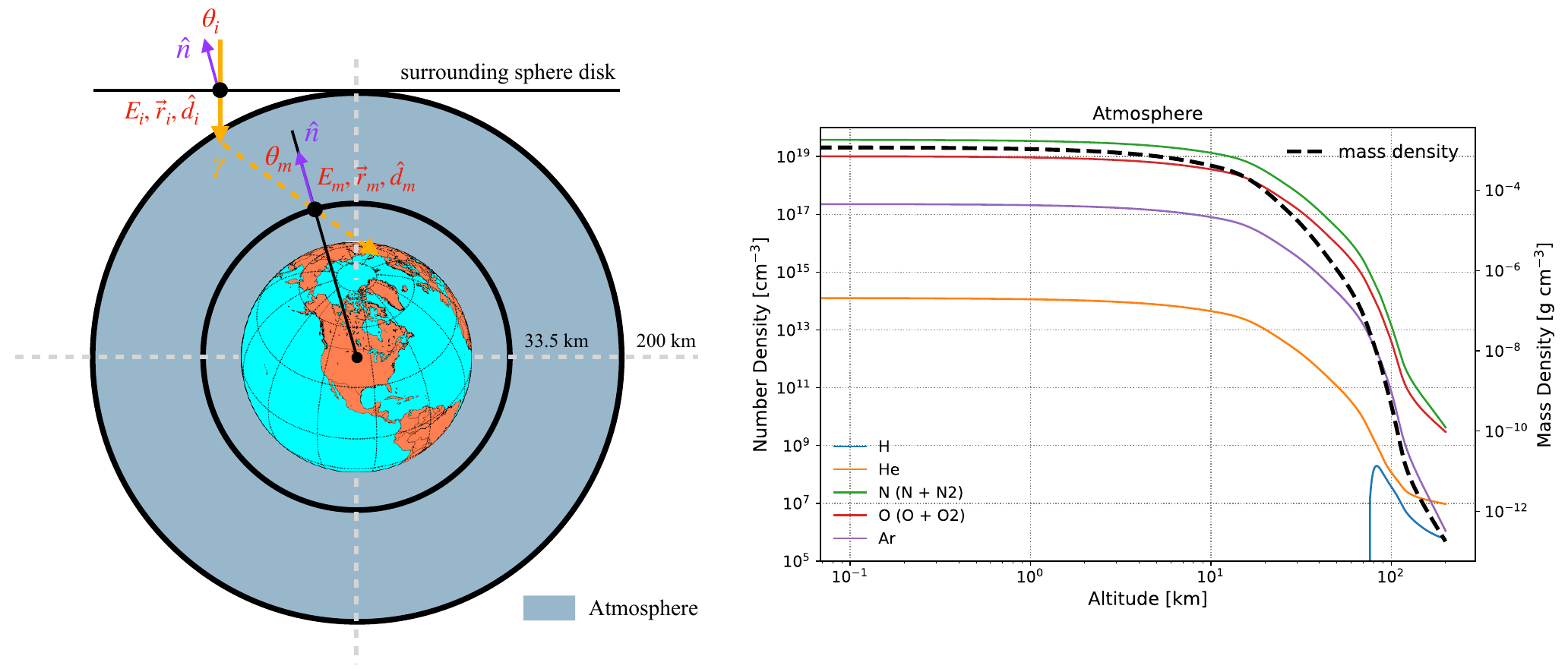}
\caption{\textbf{Left:} Schematic showing the simulation setup (not to scale). We use a spherically symmetric mass model of the atmosphere, defined relative to Earth's surface. The atmosphere is characterized using NRLMSIS, corresponding to the plot on the right. See text for more details. \textbf{Right:} Atmospheric profile defining the mass model. The left axis shows the number density for the primary species of the atmosphere, nitrogen (N), oxygen (O), argon (Ar), helium (He), and hydrogen (H). The right axis shows the total mass density.}
\label{fig:schematic}
\end{figure*}

\subsection{Atmospheric Mass Model}
The atmospheric response is simulated with the COSI atmosphere pipeline, which employs the Medium-Energy Gamma-ray Astronomy library (MEGAlib) software package\footnote{MEGAlib is available at \url{https://megalibtoolkit.com/home.html}}~\citep{2006NewAR..50..629Z}, based on Geant4 \citep{geant4_2003}. We create a spherically symmetric mass model of Earth's atmosphere, as depicted in the left panel of Figure~\ref{fig:schematic}. The model is comprised of spheriscal atmospheric shells having a thickness of 100 m, and extending from Earth's surface\footnote{We use Earth's equatorial radius, which is slightly larger than the polar radius of 6357 km.} ($R_\Earth=6378$ km) to an altitude of 200 km. The atmosphere is characterized using the latest version (v2.1) of the Naval Research Laboratory's Mass Spectrometer Incoherent Scatter Radar Model (NRLMSIS)\footnote{NRLMSIS is available at \url{https://swx-trec.com/msis}}~\citep{2002JGRA..107.1468P,2021E&SS....801321E,2022JGRA..12730896E}, implemented in the COSI atmosphere pipeline via the Python interface, pymsis\footnote{pymsis is available at \url{https://swxtrec.github.io/pymsis/}}. NRLMSIS is an empirical model of Earth’s atmosphere that describes the average observed behavior of temperature and density, from the ground to an altitude of roughly 1000 km. More specifically, the model specifies the altitude profile of the number density for the primary species of the atmosphere (i.e., nitrogen, oxygen, argon, and helium), as shown in the right panel of Figure~\ref{fig:schematic}. In general, this is dependent on location (latitude, longitude, altitude), time (year and day), and solar and geomagnetic activity levels.  

\subsection{Photon Tracking}
We simulate an isotropic source with a flat energy spectrum (i.e., constant number of photons per energy bin) between 10 keV $-$ 10 MeV, using $10^7$ photons. The source is simulated using a surrounding sphere with a radius of $R_\Earth + 200$ km. In order to mimic plane waves incident on the atmosphere, photons are emitted perpendicularly from a disk located at the surrounding sphere, and having the same radius (see left panel of Figure~\ref{fig:schematic}). With the isotropic source, photons are directed towards Earth's surface uniformly over the sky at all incident angles. The electromagnetic processes that occur as the $\gamma$ rays pass through the atmosphere are described with the Geant4 Livermore physics models, which includes Compton scattering, pair conversion (and annihilation), photoelectric absorption, Bremsstrahlung radiation, and Rayleigh scattering. Correspondingly, secondary $\gamma$ rays may also be produced from pair production and subsequent annihilation in the atmosphere, as well as from Bremsstrahlung radiation of electrons and positrons. The simulations track the $\gamma$-ray transport, including initial ($i$) and ``measured" ($m$) values of the photon's energy ($E$), position ($\vec{r}$), and direction ($\hat{d}$), as indicated in the left panel of Figure~\ref{fig:schematic}. The ``measured" values are obtained by tracking the photons whenever they cross a watched\footnote{In this context, ``watched" just refers to the volume being used to monitor the photon properties.} volume, consisting of a spherical shell at a radius of 33.5 km. With this method, our calculations are completely independent of any specific detector. It should be noted that photons in the simulations may cross the watched volume numerous times, and for each crossing the tracked information is stored. In general though, each successive crossing is increasingly unlikely, as most photons that go to lower altitudes do not back scatter, and the ones that do scatter back above the balloon altitude are highly unlikely to scatter back again. There are certainly cases where analyzing multiple crossings for a single photon becomes relevant. However, in this work we only consider the first crossing, which is a reasonable simplification considering that an Earth limb cut of $\sim90^\circ$ is typically employed in the real data analysis. 

\subsection{Determination of Incident Angles}
\label{sec:incident_angle}
From the tracked values, the incident angle ($\theta$) relative to the surface normal ($\hat{n}$) can be determined, for both initial and measured photons. The normal vector at any point on a sphere can be calculated from the position vector at that point, i.e., $ \hat{n} = \vec{r}/\lVert\vec{r}\rVert$. The incident angle is then given by the standard equation: 
\begin{equation}
    \theta = \mathrm{cos}^{-1}(-\hat{n} \cdot \hat{d}),
\label{eq:angle}
\end{equation}
where both $\hat{n}$ and $\hat{d}$ are unit vectors. The normal vector for each event is determined from the position of the measured photon, and this is also used for determining the initial incident angle. In this scheme, $\theta_i$ is undefined for photons that aren't measured. This occurs in two situations: 1) Photons are not emitted towards the watched volume, and do not get scattered into it. 2) Photons are initially emitted towards the watched volume, but get scattered away from it. While photons in the first case can be disregarded, photons in the second case must be accounted for in order to get the normalization of the response correct. The reason for this is that for any given initial incident angle, we must count the photons that get scattered and never reach the surface.

In order to determine the initial incident angle for unmeasured photons, we can find where the trajectory defined by the initial photon parameters intersects the watched volume. This will give us the normal vector that can be used in Eq.~\ref{eq:angle}. The standard vector equation of a sphere centered at the origin with radius $r$ is given by
\begin{equation}
    ||\vec{x}_s||^2 = r^2, 
\label{eq:vec_sph}
\end{equation}
where $\vec{x}_s$ are points on the sphere. The standard vector equation of a line is given by 
\begin{equation}
    \vec{x}_l = \vec{o} + D \hat{u}, 
\label{eq:vec_line}
\end{equation}
where $\vec{o}$ is the origin of the line (which in our case corresponds to $\vec{r}_i$), $D$ is the distance from the origin, $\hat{u}$ is a unit vector giving the direction of the line (corresponding to $\hat{d}_i$), and $\vec{x}_l$ gives points on the line. Plugging Eq.~\ref{eq:vec_line} into Eq.~\ref{eq:vec_sph} allows us to solve for $D$:
\begin{equation}
    D = \frac{(-\hat{u} \cdot \vec{o}) \pm \sqrt{(\hat{u} \cdot \vec{o})^2 - ||\hat{u}||^2 (||\vec{o}||^2 - r^2)}}{||\hat{u}||^2}, 
\label{eq:intersect}
\end{equation}
where we take the shortest distance because we want the first intersection. Once the intersection is known, we can determine the initial incident angle. From our $10^7$ simulated photons, $4.7\times10^5$ (4.7\%) have no (real) solution to Eq.~\ref{eq:intersect}, meaning that they were not initially directed towards the watched volume. The total number of initially thrown photons used in the response normalization is determined by this number, which gives $9.53\times10^6$. In the end, $1.5\times10^6$ photons (15\%) are not measured, which includes a majority of the photons which were not initially directed towards the watched volume, as well as photons that get scattered away from the watched volume.

\subsection{Geometric Correction Factor}
The goal of our simulations is to characterize the atmospheric response independent of any assumption on the detector geometry. Therefore, we need to apply a geometric correction factor, $\zeta_\mathrm{g}(\theta_i,\theta_m)$, to the measured photons, in order to account for the 2-dimensional detecting area used in the simulations. Such a factor has also been employed in~\cite{1995ApJ...439..963P}, and here we follow a similar approach. Because the curvature\footnote{In general, the curvature, $k$, is defined as $k = 1/r$, where $r$ is the radius of curvature.} of Earth is very small, we can approximate it as being locally flat over the general area where we expect a majority of the scattered photons to travel from. Under this assumption, for a flux, $F$, from a direction, $\theta$, the counts (N) through a horizontal patch of area ($a$) of the watched surface has a cos($\theta$) factor due to the projection:
\begin{equation}
    N = F(\theta) \times t \times a \times \mathrm{cos}(\theta),
\end{equation}
where $t$ is the exposure time. Accordingly, the ratio of measured flux ($F_m$) to initial flux ($F_i$) for respective directions $\theta_m$ and $\theta_i$ is:
\begin{equation}
\begin{split}
    \frac{F_m(\theta_m)}{F_i(\theta_i)} & = \frac{N_m}{a\mathrm{cos}(\theta_m) t} \times \frac{a\mathrm{cos}(\theta_i) t}{N_i} \\
    & = \frac{N_m}{N_i} \times \frac{\mathrm{cos}(\theta_i)}{\mathrm{cos}(\theta_m)}.
\end{split}
\end{equation}
In order to properly normalize the response, we must account for the difference in the projected area. Thus, this implies that the geometric correction factor is given by: 
\begin{equation}
   \zeta_\mathrm{g}(\theta_i,\theta_m) = \frac{\mathrm{cos}(\theta_i)}{\mathrm{cos}(\theta_m)} . 
\end{equation}
When constructing the atmospheric response matrices, the number of observed counts, $N_m(\theta_i,\theta_m)$, is weighted by this factor. We have verified that the distributions of $\vec{r}$, $\hat{d}$, $E$, and $\theta$ obtained from the simulations are consistent with the expectations from the simulation setup, as discussed in Appendix~\ref{sec:photon_distributions}.   

\section{Atmospheric Response} \label{sec:response}

\subsection{Characterizing the Response}
\label{sec:edisp}
For a given altitude ($A$), the atmospheric response ($\epsilon_\mathrm{atm}$) can be quantified in terms of initial and measured values of energy, incident angle, and azimuth angle, corresponding to the six parameters $E_i, \ E_m,\  \theta_i,\ \theta_m, \ \phi_i, \ \mathrm{and} \ \phi_m$. In this work we consider two specific representations of the response. For the first representation, we define an azimuth angle ($\phi_1$) with respect to the zenith of the detector, relative to the photon's initial position. Specifically, we use the difference between the azimuth angles for the vectors $\hat{d_m}$ and $\hat{d_i}$, which in spherical coordinates can be expressed as 
\begin{equation}
    \phi_1 = \phi_{dm} - \phi_{di},
\end{equation}
where $\phi_{dm}$ is the azimuth of $\hat{d_m}$, and $\phi_{di}$ is the azimuth of $\hat{d_i}$. Thus, for this first case, the response is represented in terms of $E_i, \ E_m,\  \theta_i,\ \theta_m, \ \mathrm{and} \ \phi_1$. This representation is particularly useful when dealing with diffuse sources, such as the Galactic diffuse or extragalactic gamma-ray background. In these cases, photons enter the detector from all directions, and typically we can simplify the response by summing over all azimuth angles. Indeed, one of the primary motivations for this work is to quantify the atmospheric response for the scattered component, which is most important when analyzing diffuse sources. This representation is also applicable when dealing with non-imaging telescopes. 

For the second representation, we characterize the response with respect to the position of the source on the sky. To accomplish this, we use the difference of incident directions:
\begin{equation}
\Delta \theta_\mathrm{s} = \mathrm{cos}^{-1}(\hat{d_i} \cdot \hat{d_m}). 
\end{equation}
Additionally, we define an azimuth angle ($\phi_2$) in the perpendicular plane to the source position. The normal to the plane is given by $\hat{d}_i$, and the initial point in the plane is obtained by the orthogonal component ($\hat{z}_\perp$) of the projection of $\hat{z}$ onto $\hat{d_i}$, where $\hat{z}$ is the zenith of the instrument. For the initial direction ($\hat{\phi}_2$) we use the cross product:
\begin{equation}
   \hat{\phi}_2 = \hat{d_i} \times \hat{z}_\perp.
\end{equation}
Thus, for this second case, the response is represented in terms of $E_i, \ E_m,\  \theta_i,\ \Delta \theta_\mathrm{s}, \ \mathrm{and} \ \phi_2$. This representation is useful for analyzing point sources, observed with an imaging telescope, such as COSI. In this case, photons can be selected coming from the direction of the source, which is primarily limited by the angular resolution of the instrument. This is discussed further in Section~\ref{sec:imaging_telescopes}. 

\subsection{Convolving the Response}
\label{sec:convolve}
\begin{figure*}
\centering
\includegraphics[width=2\columnwidth]{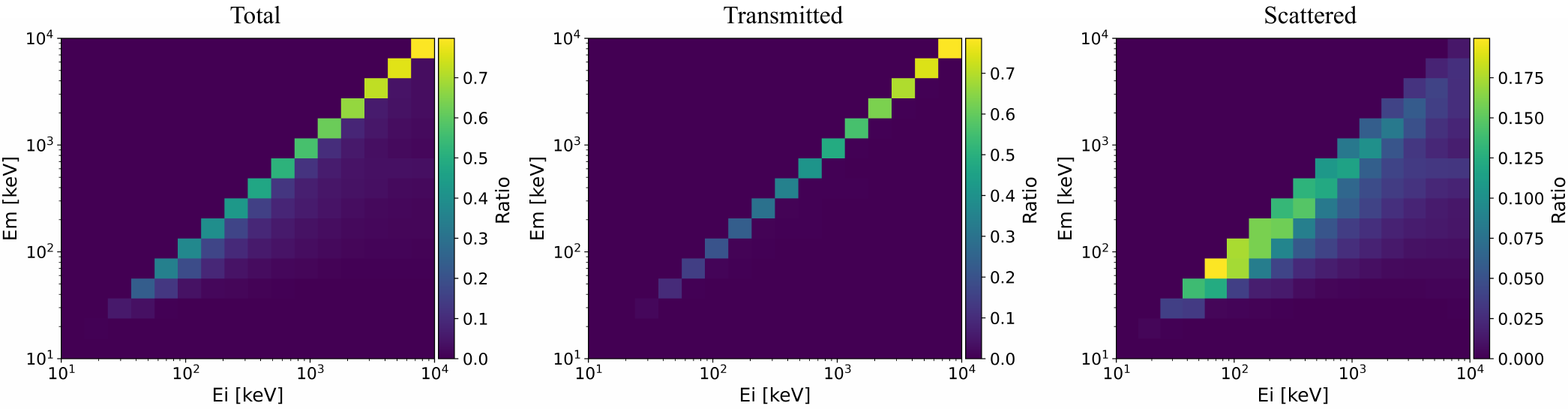}
\caption{Energy dispersion matrix for all photons (left), transmitted photons (middle), and scattered photons (right), for a 50$^\circ$ off-axis source. The y-axis is the measured photon energy, and the x-axis is the initial photon energy. The matrices are normalized by the total photons simulated in each bin of $E_i$ and $\theta_i$. The total energy dispersion matrix is the sum of the transmitted and scattered components (note that the components are shown with different colorbar ranges).}
\label{fig:energy_dispersion_50deg_sphere}
\end{figure*}
In order to completely account for the atmospheric response, a full convolution with the detector response ($R$) must be made. Considering our first representation of the atmospheric response at a given altitude, for an off-axis angle $\theta_i$\footnote{In this work we consider an instrument with a zenith pointing, in which case ``off-axis" and ``zenith" are synonymous.} and an exposure time $\Delta t$, the predicted counts in each energy bin can be calculated as 
\begin{equation}
\begin{split}
    P = \Delta t \iiint_{\phi_1, \theta_m, \Delta E}
 R(Z,\phi,E_i,E_m) \\
\circledast \epsilon_\mathrm{atm} (A,\theta_i,\theta_m,E_i,E_m,\phi_1) 
 \circledast F(E,\vec{\alpha}) \\ \ dE \ d\theta_m \ d\phi_1,
\end{split}
\end{equation}
where $Z$ and $\phi$ are the zenith and azimuth of the photon source, respectively, $F$ is the input spectral model, and $\vec{\alpha}$ gives the model parameters of the spectrum. Note that the convolution must map $\theta_m$ to $Z$ and $\phi_1$ to $\phi$, in the working reference frame. This convolution fully accounts for the fact that the scattered photons are detected at different off-axis angles compared to the source, and thus detected with a different detector response. A similar convolution would be made for our second representation of the atmospheric response, except that we would need to replace $\phi_1 \rightarrow \phi_2$ and $\theta_m \rightarrow \Delta \theta$.

\subsection{Energy Dispersion Matrices}  \label{sec:edisp}
In order to examine some basic properties of the atmospheric response for both the transmitted and scattered components, in this section we calculate energy dispersion matrices. As our representative case we consider a source with an off-axis angle of $50^\circ$, and we project the response onto the axes $E_i$ and $E_m$. We bin the simulations using $4^\circ$ angular bins and 16 log-spaced energy bins ($\sim$ 5.33 bins per decade). This particular binning is primarily motivated by the need to obtain sufficient statistics in each bin. Additional photons can always be simulated if finer bins are required, such as for the detailed analysis of spectral lines. Note that the angular binning is comparable to the resolution of the COSI balloon instrument: $5.9^\circ$ at 0.511 MeV; $3.9^\circ$ at 1.809 MeV~\citep{BeechertCOSICalib}.  

Figure~\ref{fig:energy_dispersion_50deg_sphere} shows the energy dispersion matrices for both the scattered and transmitted photons, as well as the sum of the two components. We identify a photon as having undergone at least one scatter if its measured incident angle varies from its initial incident angle. Of course with this approach we are limited by the angular resolution of our incident angle bins, i.e., we cannot resolve a photon that scatters within $4^\circ$. The matrices are normalized by the total number of photons simulated in each bin of $E_i$ and $\theta_i$, as discussed in Section~\ref{sec:incident_angle}. The color scale gives the ratio of photons that are detected with a measured energy, $E_m$, given an initial energy, $E_i$. For the transmitted component, the ratio is a proper probability. However, for the scattered component, the measured incident angle is different than the initial incident angle, and thus the ratio is not a true probability (as it is not normalized to unity).  

The transmitted photons are shown in the middle panel of Figure~\ref{fig:energy_dispersion_50deg_sphere}. As can be seen, this component resides completely along the main diagonal, with a probability that steadily increases towards the upper energy bound. The photons are along the main diagonal because they are the ones that reach the detector without scattering, and therefore there is no energy loss.  

The scattered photons are shown in the right panel of Figure~\ref{fig:energy_dispersion_50deg_sphere}. The distribution of the scattered component is substantially different than that of the transmitted component. Most notably, there is a significant number of off-diagonal photons, and the ratio is highest towards lower energies. A diagonal component is still present, but this is mostly an artifact of the coarse energy binning. Based on standard Compton dynamics, the photon's energy after a scattering ($E'$) is a function of the scattering angle $\theta$:
\begin{equation}
E' = \frac{E_i}{1+\frac{E_i}{m_e c^2}(1-\cos \theta)}~,
\label{eq:comptondynamics}
\end{equation}
where $m_e c^2 \simeq 0.511\,\mathrm{MeV}$ is the rest-mass energy of the electron. For $E_i / {m_e c^2} \ll 1$, $E' \sim E_i$ for all values of $\theta$, leading to all scattered low-energy photons piling up close to the diagonal. The scattered component also has a prominent feature in the energy bin containing the 0.511 MeV line. This is due to pair production and subsequent annihilation in the atmosphere. The electron-positron pairs are produced from $\gamma$-ray photons interacting with the Coulomb field of electrons or nuclei in the atmosphere. More details regarding the photon interaction sequences that occur in the simulations are given in Appendix~\ref{sec:photon_interactions}.

\begin{figure}
\centering
\includegraphics[width=0.99\columnwidth]{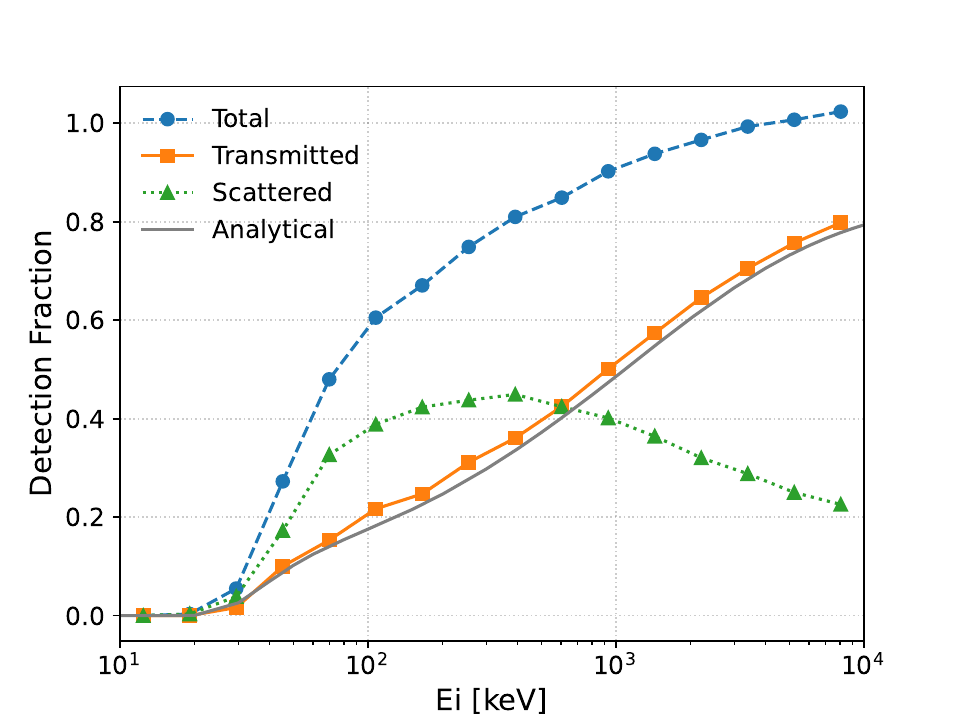}
\caption{Detection fraction for the three components, as specified in the legend, for a 50$^\circ$ off-axis source. The detection fraction is the projection of the energy dispersion matrix (see Figure~\ref{fig:energy_dispersion_50deg_sphere}) onto the initial energy axis, and for the transmitted component it is equivalent to the transmission probability.}
\label{fig:tp_50_deg}
\end{figure}

The detection fraction can be obtained by projecting the energy dispersion matrix onto the initial energy axis, as shown in Figure~\ref{fig:tp_50_deg}. Note that for the transmitted component, the detection fraction is equivalent to the TP. These results clearly show that the effect from the scattered component is most dominant for initial energies between $\sim0.2 - 0.4$ MeV. Moreover, the detection fraction of the scattered component exceeds that of the transmitted component for energies below $\sim$ 0.6 MeV. For comparison, we also show the TP for the transmitted component calculated analytically. As can be seen, the analytical calculation is in very good agreement with the simulations, which is another important validation of our simulation pipeline. We note that the results from the simulation appear to be slightly higher than the analytical result, but this is due to the coarse binning in angle and energy (note that data points are plotted at the geometric mean). Details of the analytical calculation are provided in Appendix~\ref{sec:tp_analytical}.

\subsection{Approximating the Atmospheric Response} \label{sec:applications}

\begin{figure*}[t]
\centering
\includegraphics[width=0.95\columnwidth]{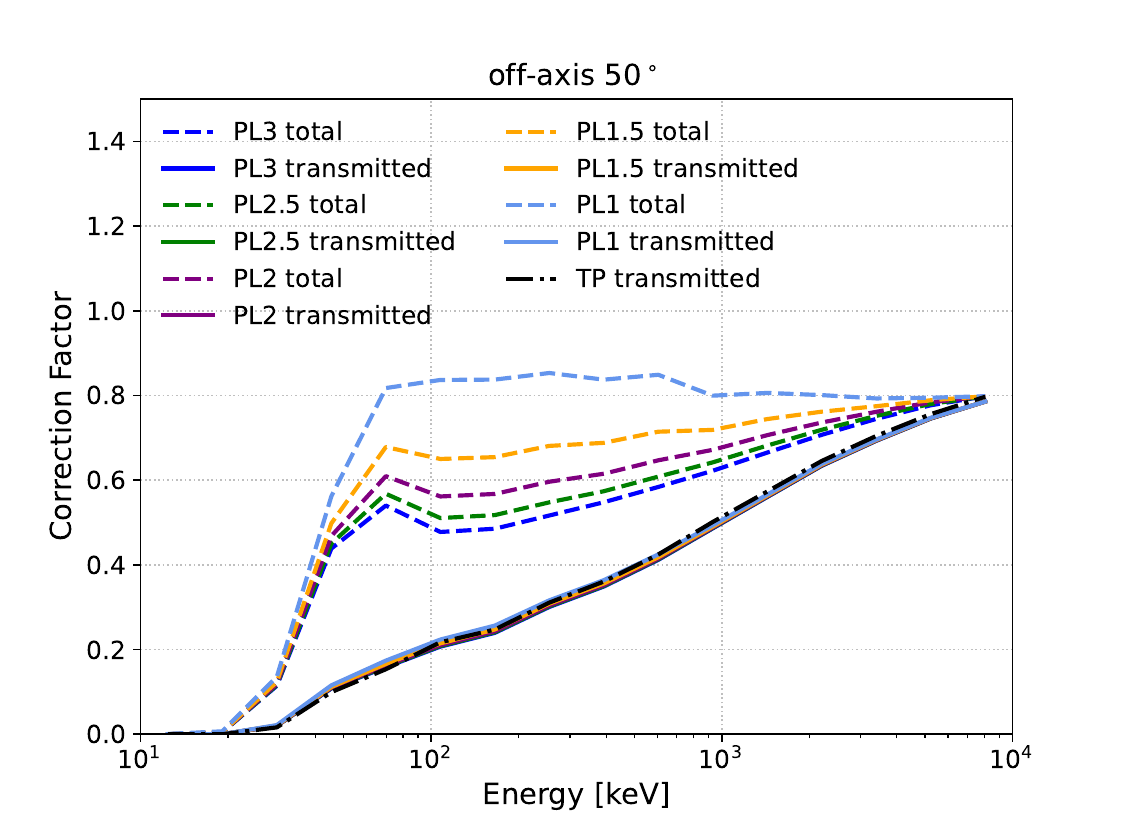}
\includegraphics[width=0.95\columnwidth]{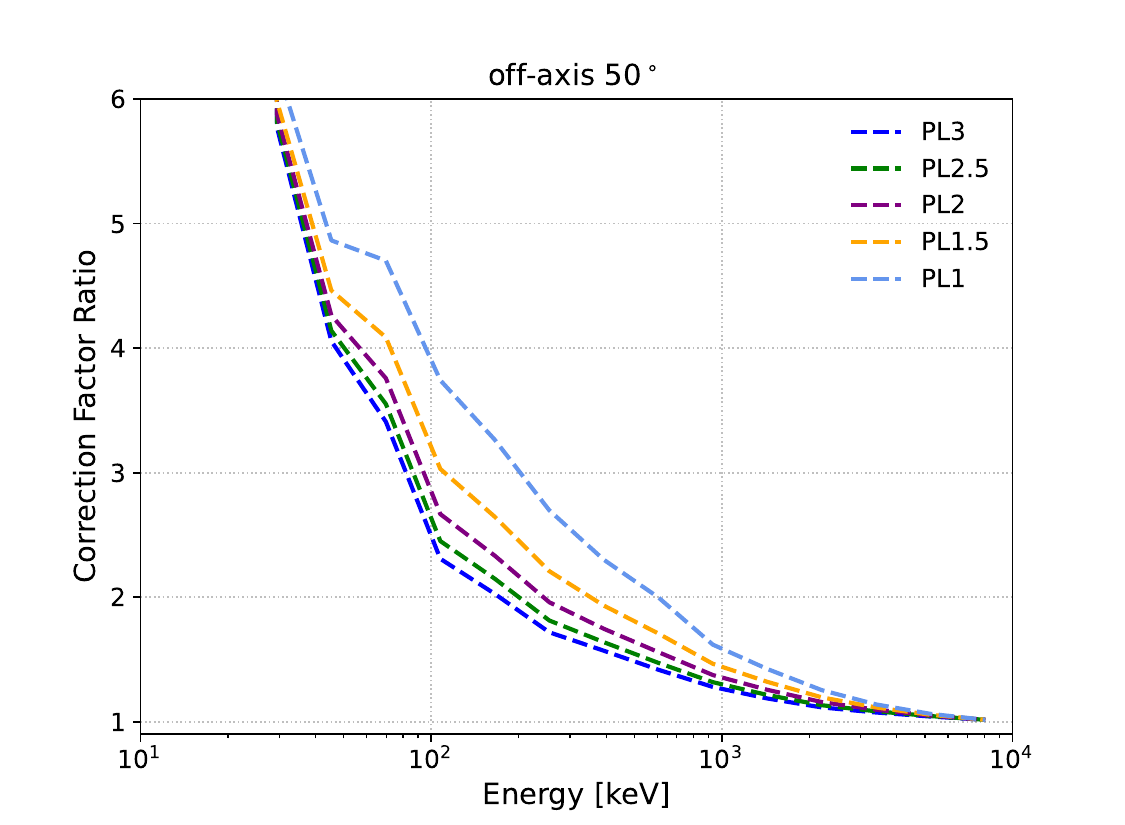}
\caption{\textbf{Left:} Correction factor (as defined in Eq.~\ref{eq:correction}) for a range of power law spectral models ranging from $1-3$, for a 50$^\circ$ off-axis source. The solid line is for the transmitted component only (note that the models mostly overlap), and the dashed line is for the total component, which includes transmitted and scattered photons. The black dash-dot line is the TP, obtained by projecting the energy dispersion matrix onto the initial energy axis. \textbf{Right:} Correction factor ratio (as defined in Eq.~\ref{eq:correction_ratio}) for the same range of models shown in the left plot.}
\label{fig:correction_sphere_50}
\end{figure*}
As discussed in Section~\ref{sec:convolve}, the proper way to correct for atmospheric effects is to start with a full convolution of the atmospheric response with the detector response. However, this requires detailed aspect information of the instrument during the observations, and in general the calculation can be complex. It would therefore be helpful to have a simplified approach that could be used to estimate atmospheric effects. To this end, here we calculate the ratio of the predicted counts to the model counts, for a given spectral input, which we refer to as the correction factor. The predicted counts are obtained by forward-folding the atmospheric response with the input spectrum, where we use the energy dispersion matrices discussed in the previous section. Thus, in this case we are assuming that the source is at an average off-axis angle, and we are integrating over all azimuth angles. Moreover, we are neglecting the fact that the scattered photons will be measured at different incident angles, and thus with a different detector response. Despite these simplifications, this approach provides a reasonable approximation of the atmospheric effects for diffuse sources, such as the Galactic diffuse continuum emission~\citep{COSI:2023bsm}. 

As a toy example, we can consider a source with three generic energy bins. For a given altitude and off-axis angle, the forward-folding is obtained as follows:
\begin{equation}
    \epsilon_\mathrm{atm}\vec{F} = \vec{P},
\end{equation}
where $\epsilon_\mathrm{atm}$ is the energy dispersion matrix (as shown in Figure~\ref{fig:energy_dispersion_50deg_sphere}), $\vec{F}$ is the model flux (i.e., the source flux incident on the atmosphere), and $\vec{P}$ is the predicted flux (i.e., the resulting flux incident on the instrument after accounting for atmospheric effects), both in units of $\mathrm{ph \ cm^{-2} \ s^{-1}}$. Symbolically, the matrix multiplication can be written out as follows: 

\begin{equation*}
    \begin{bmatrix}
    0 & 0 & \epsilon_{33} \\
    0 & \epsilon_{22} & \epsilon_{23}\\
    \epsilon_{11} & \epsilon_{12} & \epsilon_{13}
    \end{bmatrix} 
    \begin{pmatrix}
    F_1 \\F_2\\F_3
    \end{pmatrix}
    = 
     \begin{pmatrix}
    P_3 \\
    P_2\\
    P_1
    \end{pmatrix},
\end{equation*}
where the indices specify the energy channel. For the energy dispersion matrix, the first index gives the row (corresponding to the measured energy axis), and the second index gives the column (corresponding to the initial energy axis). Note that the upper triangle of the energy dispersion matrix is always zero, as the off-diagonal entries are found in the lower triangle, due to photons which lose energy after scattering. From this we obtain
\begin{flalign*}
    P_1 =& \ \epsilon_{11}F_1 + \epsilon_{12}F_2 + \epsilon_{13}F_3  \\
P_2 =& \ \epsilon_{22}F_2 + \epsilon_{23}F_3 \ \\ 
P_3 =& \ \epsilon_{33}F_3
\end{flalign*}
Writing out the equations in this way is helpful for gaining insight into the dynamics of how the scattered component can produce an energy-dependent distortion in the predicted spectrum. For each energy channel the predicted counts is the sum of the flux from all channels at and above the given energy, weighted by the respective ratios. In the case of the transmitted component, the off-diagonal entries are all zero, and the predicted flux in a given energy bin is a fraction of the model flux in that same bin (corresponding to the TP). However, for the scattered component, the off-diagonal entries are not all zero, and thus the predicted counts in a given energy bin will have contributions from higher energy bins.   

The energy-dependent correction factor, $c$, for a given energy bin, $E_i$, can now be defined as
\begin{equation}\label{eq:correction}
    c(E_i) =  P_i/F_i.
\end{equation}
The left panel of Figure~\ref{fig:correction_sphere_50} shows the correction factor for a range of power law spectral models with different assumptions on the spectral index. The correction factor is shown for both the transmitted component, as well as the total component. For the former, the correction factor is equivalent to the TP, and thus it is the same for all spectral indices. On the other hand, the total correction factor has a dependence on the spectral index, due to the scattered photons. Harder sources will generally have more photons at higher energies that get scattered down to lower energies. Indeed, this is the exact trend that we find. 

As shown by the correction factor, the scattered component can have a significant impact on the atmospheric response, compared to including only the transmitted component. In order to quantify this further, we define the correction factor ratio, $R(E_i)$, as
\begin{equation}\label{eq:correction_ratio}
    R(E_i) = \frac{c_\mathrm{tot}(E_i)}{c_\mathrm{tran}(E_i)},
\end{equation}
where $c_\mathrm{tot}$ and $c_\mathrm{tran}$ are the correction factors for the total and transmitted components, respectively. The right panel of Figure~\ref{fig:correction_sphere_50} shows the correction factor ratio for the same range of power law spectral models shown in the left panel. As can be seen, the scattered component is most important towards lower energies. Note that photoelectric absorption becomes dominant for energies below $\sim$ 100 keV, and the atmosphere rapidly becomes opaque to $\gamma$-ray photons. In this regime, the simulations have very low statistics, and the correction factor ratio becomes very large. In this work we are mainly interested in energies $>0.1$ MeV, and lower energies will be the focus of a future study.  

For balloon-borne observations of a diffuse source at a given altitude, Eq.~\ref{eq:correction} can be applied to correct for the atmospheric response. This would be equivalent to scaling the effective area. Alternatively, if the transmitted component has already been accounted for (i.e., in the determination of the effective area), Eq.~\ref{eq:correction_ratio} can be applied to correct for the scattered component. However, we again stress that these corrections only provide approximations, and in order to completely correct for the atmospheric response, a full convolution with the detector response must be made, as described in Section~\ref{sec:convolve}. 

So far we have only considered an off-axis angle of $50^\circ$. More generally, a similar trend is found for all off-axis angles. For higher off-axis angles, the detection fraction for both the transmitted and scattered components decreases, although the correction factor ratio increases, essentially due to the fact that there is more atmosphere for the photons to traverse. As another example, in Appendix~\ref{sec:on_axis_spherical} we show the results for an on-axis source. 

\subsection{Point Sources and Imaging Telescopes}
\label{sec:imaging_telescopes}

\begin{figure}[t] 
\centering
\includegraphics[width=0.9\columnwidth]{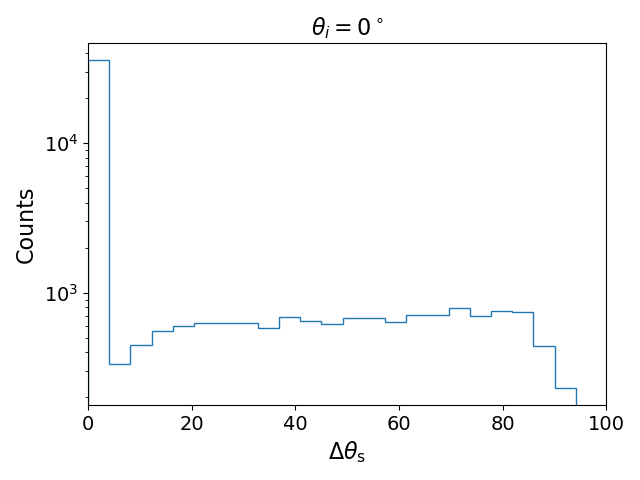}
\includegraphics[width=0.94\columnwidth]{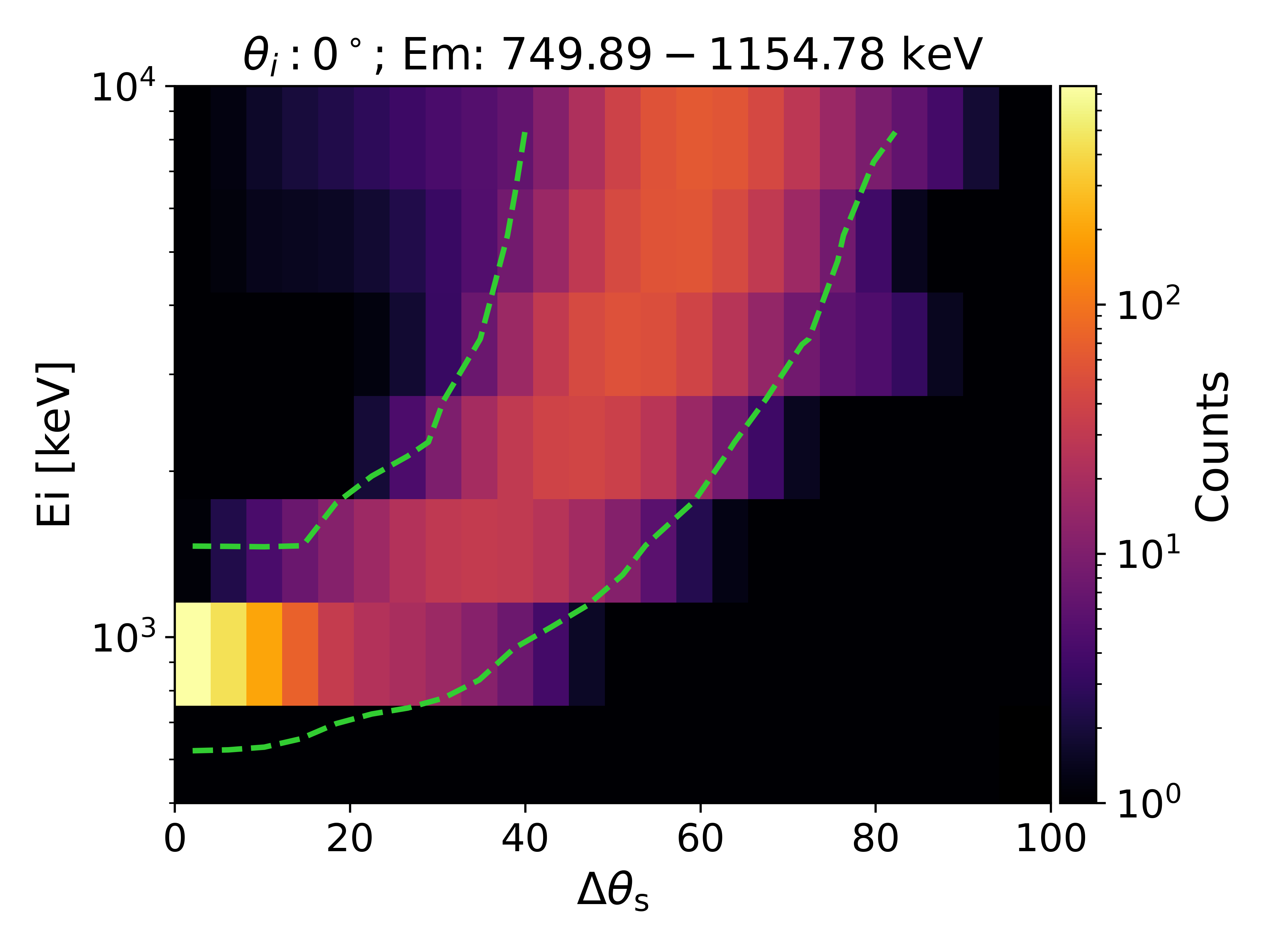}
\includegraphics[width=0.9\columnwidth]{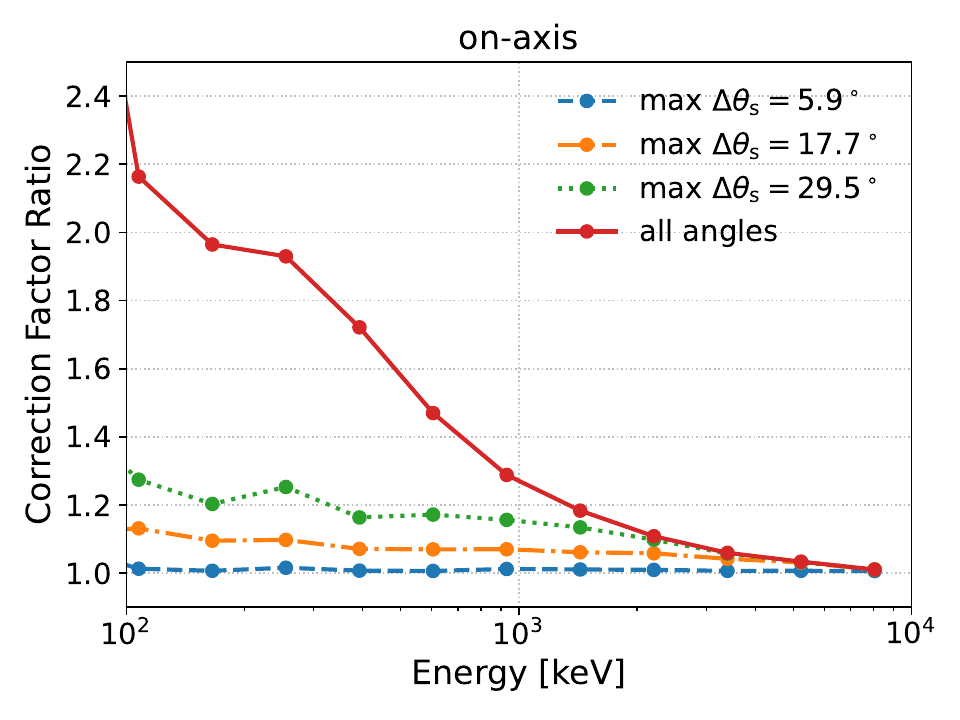}
\caption{\textbf{Top:} Distribution of $\Delta \theta_\mathrm{s}$ for an on-axis source. \textbf{Middle:} 2-dimensional distribution of $\Delta \theta_\mathrm{s}$ and initial energies, for measured energies between $0.750 - 1.155$ MeV. The image is smoothed with a 2D Gaussian kernel, with standard deviations of 1 and 0.1 (pixels) in the x and y directions, respectively. The dashed green contour is at the level of the max array value divided by 65. \textbf{Bottom:} Correction factor ratio calculated for different max values of $\Delta \theta_\mathrm{s}$, as specified in the legend.}
\label{fig:rep2_plots}
\end{figure}

For point sources observed with an imaging telescope, such as COSI, Eqs.~\ref{eq:correction} and~\ref{eq:correction_ratio} can also be applied as an approximation of the atmospheric response. However, in this case, photons can be selected coming from the direction of the source, mainly limited by the angular resolution of the instrument. The approximation should therefore be made using the alternative representation of the atmospheric response (in terms of $E_i, \ E_m,\  \theta_i,\ \Delta \theta_\mathrm{s}, \ \mathrm{and} \ \phi_2$). This allows us to calculate the energy dispersion matrices using a slice of $\Delta \theta_\mathrm{s}$, where we can choose a max value corresponding to the instrument's angular resolution.

As a specific example, here we consider an on-axis source ($\theta_i=0^\circ$), and we use the angular resolution of the COSI balloon instrument ($5.9^\circ$ at 0.511 MeV). The top panel of Figure~\ref{fig:rep2_plots} shows the distribution of $\Delta \theta_s$. As can be seen, a majority of the photons are in the first bin, corresponding to the transmitted component. Additionally, the distribution shows a sharp cutoff at $90^\circ$, due to the fact that we are only considering the first time a photon crosses the watched volume. The middle panel of Figure~\ref{fig:rep2_plots} shows the 2-dimensional distribution of $\Delta \theta_\mathrm{s}$ and initial energies, for measured energies between $0.750 - 1.155$ MeV. Quite interestingly, this plot reveals a characteristic property of the scattering; that is, as we observe at higher angular distances from the source location, a majority of the measured photons in a given energy band originate at increasingly higher energies. Finally, the bottom panel of Figure~\ref{fig:rep2_plots} shows the correction factor ratio calculated for different max values of $\Delta \theta_\mathrm{s}$, where we consider 1$\times$, 3$\times$, and 5$\times$ the angular resolution of the COSI balloon instrument at 0.511 MeV ($5.9^\circ$, $17.7^\circ$, and $29.5^\circ$, respectively). The calculations are made using a power law spectral model with a photon index of 2.0. As can be seen, the ability to select photons within a limited angular distance of the source position substantially reduces the effects of scattering. In terms of the correction factor ratio, the effect is $\lesssim 30\%$ when including up to 5$\times$ the angular resolution.  

As a check on our calculations, we have verified that for the second representation of the atmospheric response used here, we obtain consistent results for the detection fraction (i.e., Figure~\ref{fig:TP_on_axis_sphere}) in the limiting case of including all angles, for both transmitted and scattered components. Indeed, the correction factor ratio for all angles in Figure~\ref{fig:rep2_plots} is consistent with the results obtained with the other representation of the atmospheric response.

\subsection{Validation of the Simulations}

\begin{figure}[t]
\centering
\includegraphics[width=0.95\columnwidth]{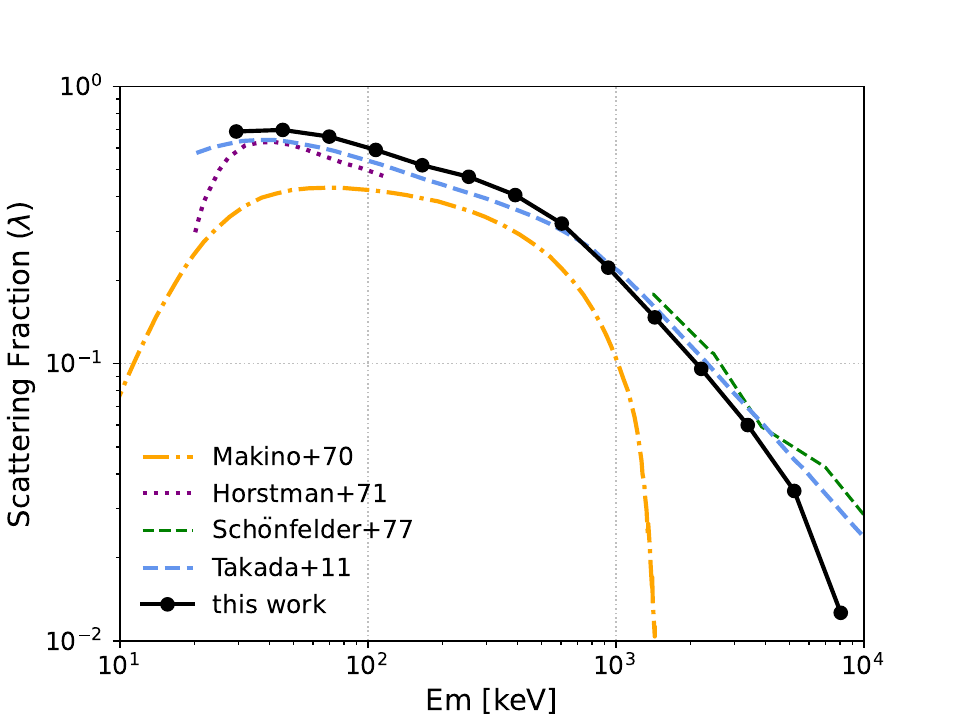}
\caption{Scattering fraction as a function of measured photon energy. Results from this work are shown in black, which are compared to a number of other calculations/simulations from the literature, as specified in the legend.}
\label{fig:sfrac}
\end{figure}

As an additional check on our simulations, we compare our results to estimates of the
atmospheric response from~\citet{2011ApJ...733...13T}, which was used to successfully fit the
growth curve of the SMILE balloon flight. To do this we calculate the scattering ratio,
defined as
\begin{equation}
    \lambda(a,Z) = \frac{F_s(a,Z)}{F_t(a,Z)+F_s(a,Z)},
\end{equation}
where $F_t(a,Z)$ and $F_s(a,Z)$ are the flux of the transmitted and scattered components, respectively, for an atmospheric depth, $a$, and zenith angle, $Z$. We use an atmospheric depth of $7.8 \ \mathrm{g \ cm^{-2}}$, corresponding to an altitude of 33.5 km and temperature of 234.35 K. We assume a power law spectral model with an index of 2, and consider zenith angles between $0^\circ - 20^\circ$. These are essentially the same selections that were used in~\citet{2011ApJ...733...13T}, with the exception that their atmospheric depth was actually $8.0 \ \mathrm{g \ cm^{-2}}$. Figure~\ref{fig:sfrac} shows the comparison of the scattering fraction. The figure also includes a number of other estimates from the literature~\citep[see][and references therein]{2011ApJ...733...13T}. Note that all literature results are from simulations or calculations. Overall, our results are in very good agreement with the results from~\citet{2011ApJ...733...13T}. We note that near 10 MeV our results are slightly lower, but this is also the upper energy bound of our analysis, and thus the comparison here is not very reliable.

As another sanity check on our simulations we have analyzed the atmospheric response using a rectangular mass model with a narrow beam as the $\gamma$-ray source. In fact, although unknown to us when originally developing the simulations, a similar approach was actually taken in~\citet{1995ApJ...439..963P} for balloon-borne measurements of the supernova SN 1987A in the hard X-ray continuum. One of the main benefits of the rectangular mass model is that it can provide a somewhat simpler and intuitive approach to studying the atmospheric response, and it is less computationally intensive. Overall, we find that the results from the rectangular mass model are consistent with those from the spherical mass model. More details for this are provided in Appendix~\ref{sec:rectangular_mass_model}.

We have also successfully applied our atmospheric corrections to data from the 2016 COSI balloon flight. Detection of the Galactic diffuse continuum emission during the flight was recently reported in~\citet{COSI:2023bsm}. In that case, application of the correction factor ratio was found to bring the measured flux in closer agreement with previous measurements, as would be expected.  

In summary, the simulations presented in this work have been validated in the following ways:

\begin{enumerate}

   \item We have verified that the photon distributions from the simulations are in accordance with expectations from the simulation setup.  
    \item The TP calculated from simulations is in excellent agreement with the analytical calculation. 
    \item The scattering ratio is in very good agreement with other calculations from the literature. 
    \item The results from the spherical mass model were shown to be in good agreement with results from the simplified rectangular mass model. 
    \item The atmospheric corrections were successfully applied to real observational data in~\citet{COSI:2023bsm}.  
    
\end{enumerate}

\section{Summary and Conclusion} \label{sec:conclusions}

In this work we have simulated the full response for $\gamma$-ray transport in the atmosphere. The simulations are run with the COSI atmosphere pipeline, which employs MEGAlib, based on Geant4. The atmosphere is characterized using the latest version of NRLMSIS. We characterized the transport of $\gamma$ rays in the atmosphere in terms of two components. The first component is due to photons that scatter and never reach the detector, thereby causing an attenuation of the original signal. The second component is from photons that reach the detector after scattering one or more times.   

Using the energy dispersion matrices from our simulations, we have calculated the detection fractions for both photon components. Additionally, we have defined the correction factor and correction factor ratios, which approximate the effects of the scattered component on the measured spectrum, i.e., the extent to which the scattered photons cause an energy-dependent distortion. For the transmitted component, the detection fraction is equivalent to the TP, and it gives the probability that a photon will reach the detector. Since the photons that reach the detector do not undergo any scattering, they arrive at the detector with the same energy and direction as they started with. The detection fraction for the scattered component is analogous to that of the transmitted component. The main difference is that the photons arrive at the detector with different energies and directions compared to their starting values.  

Accounting for the scattered component is most important for diffuse sources because photons enter the detector from all directions. We find that the detection fraction for the scattered photons is highest for initial energies between $\sim 0.2 - 0.4$ MeV. In general, the contribution from the scattered component depends on the photon index of the source. Harder sources have more photons at higher energies, which lose energy as they are scattered. Thus, the scattered component is more dominant for harder sources. The end result is an energy-dependent spectral distortion, which is highest towards lower energies. At 0.1 MeV the scattered component may increase the flux (with respect to only accounting for attenuation) by as much as a factor of $\sim2-4$, depending on the photon index and off-axis angle of the source.

For point sources observed with imaging telescopes, such as COSI, the effect from scattering is not as important because they have the ability to decipher the direction of the incident photons, mainly limited by the angular resolution of the instrument. When including photons out to an angular distance of $29.5^\circ$ (5$\times$ the angular resolution of the COSI balloon instrument at 0.511 MeV), the effect from scattering is $\lesssim 30\%$, in terms of the correction factor ratio.
 
These results highlight the importance of accounting for photons that scatter into the detector when making balloon-borne observations, especially for diffuse sources. The simulation and analysis pipeline described in this work is publicly available and is readily applicable to observations of MeV $\gamma$ rays in the atmosphere.  

\section*{Acknowledgements}
The COSI balloon program was supported through NASA APRA grants NNX14AC81G and 80NSSC19K1389. We also acknowledge support for this work under NASA APRA grant 80NSSC21K1815. This work is partially supported under NASA contract 80GSFC21C0059, and it is also supported in part by the Centre National d’Etudes Spatiales (CNES). CMK's research was supported by an appointment to the NASA Postdoctoral Program at NASA Goddard Space Flight Center, administered by Oak Ridge Associated Universities under contract with NASA. CMK is pleased to acknowledge conversations with Alex Moiseev which helped to improve the quality of the analysis. 

\appendix 
\section{Photon Distributions}
\label{sec:photon_distributions}
\begin{figure*}[t]
\centering
\includegraphics[width=1\columnwidth]{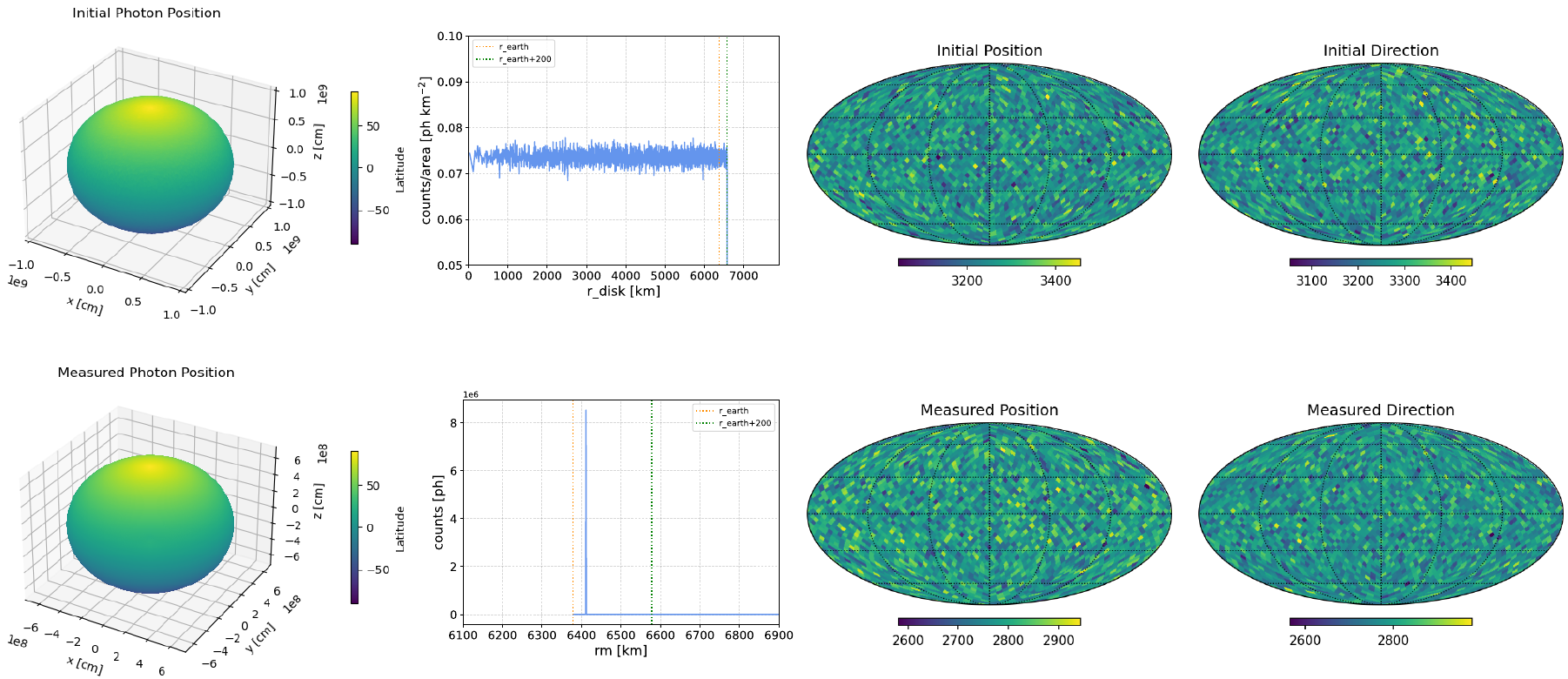}
\caption{The top row shows initial photons and the bottom row shows measured photons. The columns are as follows: Column 1: 3-dimensional scatter plot of photon position, where the color scale shows the latitude. Column 2 top: Radial distribution for the initial photons, where we show the rate (photons per area), with respect to the radius of the surrounding sphere disk. Column 2 bottom: Radial distribution for the measured photons, showing the counts, where the radius is defined relative to Earth's center (i.e., the origin of the coordinate system). The dotted orange line corresponds to Earth's radius, and the dotted green line corresponds to the altitude of the surrounding sphere. Column 3: HEALPix map showing the position on the sky. Column 4: HEALPix map showing the direction on the sky (i.e., where the direction vector is pointing).}
\label{fig:dist_plot}
\end{figure*}

In Figure~\ref{fig:dist_plot} we show the photon distributions of position and direction from the simulation, where the first row is for the initial photons, and the second row is for the measured photons. Column 1 shows a 3-dimensional scatter plot of the photon position. The photons are distributed over a spherical region, as expected. The second column shows the radial distribution. For the initial photons we show the count rate, with respect to the radius of the surrounding sphere disk. The rate is constant with a mean value of $0.074 \mathrm{\ ph \ km^{-2}}$, and goes to zero exactly at the radius of the surrounding sphere ($r_\mathrm{sphere} = R_\Earth + 200$ km). This is exactly as expected. For the measured photons the radius (with respect to Earth's center) is exactly at the location of the watched volume (6411.6 km). Columns 3 and 4 show all-sky HEALPix maps of position and direction, respectively. We use an NSIDE of 16, corresponding to an approximate angular resolution of $3.7^\circ$. These distributions show that the position and direction are uniformly distributed over the sky. Moreover, with this angular resolution we would expect $\sim 3300$ (initial) photons per pixel, consistent with what is shown. 

\begin{figure*}[t]
\centering
\includegraphics[width=0.45\columnwidth]{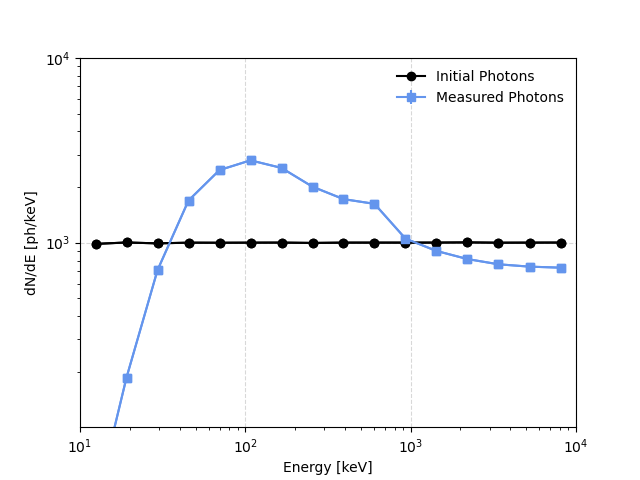}
\caption{Distributions of energy for both initial and measured photons.}
\label{fig:energy_dist}
\end{figure*}

The distribution of energies for both initial and measured photons is shown in Figure~\ref{fig:energy_dist}. For initial photons, the distribution is flat across all energies, in accordance with the simulated spectrum. The distribution of measured photons shows more photons at lower energies, which is a consequence of the energy loss from scattering. Note that in total there are $10^7$ initial photons and $8.5 \times 10^6$ measured photons.    

\begin{figure*}[t]
\centering
\includegraphics[width=0.45\columnwidth]{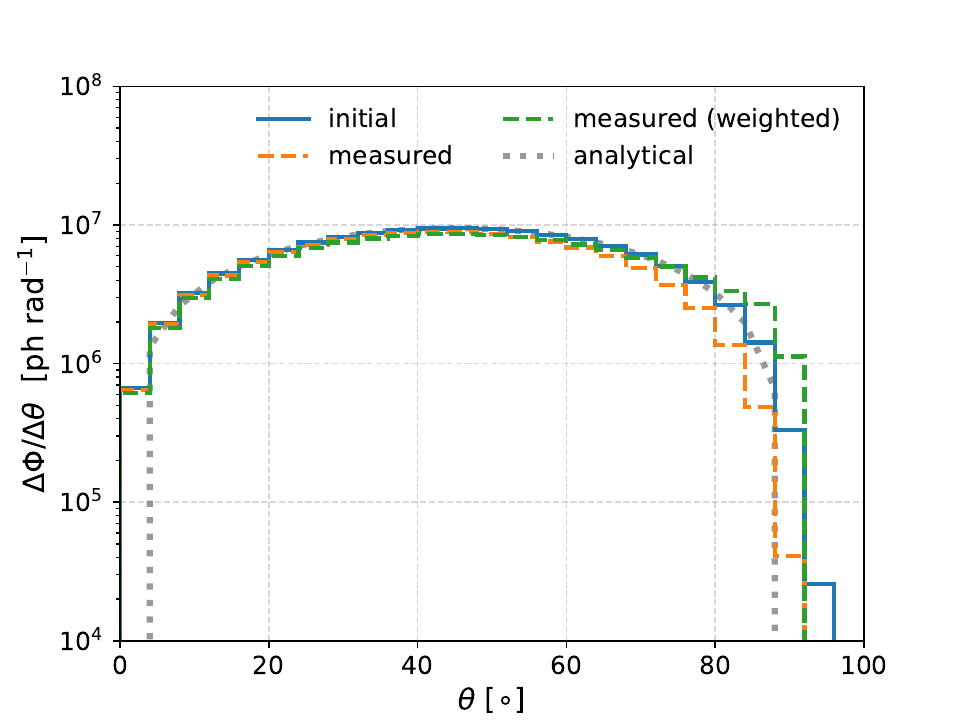}
\includegraphics[width=0.45\columnwidth]{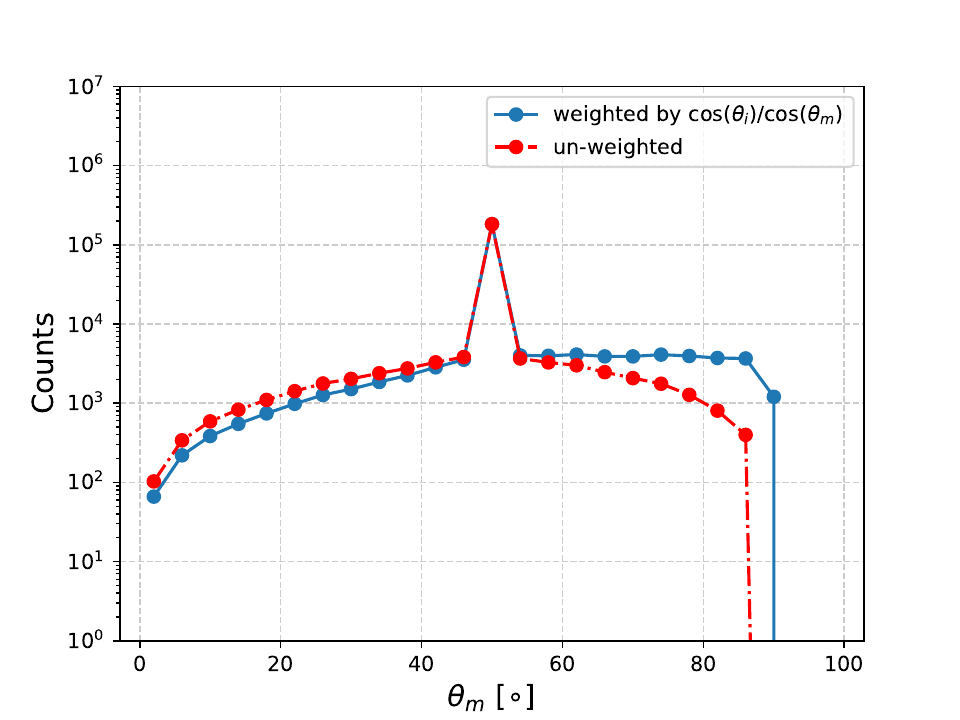}
\caption{\textbf{Left:} Distributions of incident angles for both measured (un-weighted and weighted by the geometric correction factor) and initial photons, integrated over all energies. The dashed grey curve shows the analytical solution for the flux of a constant field through a hemisphere, with input parameters corresponding to our simulation setup. \textbf{Right:} Distribution of measured incident angles, for an initial angle of $50^\circ$ and energy of 8 MeV. For comparison, we show both the weighted and un-weighted histograms.}
\label{fig:theta_dist}
\end{figure*}

The left panel of Figure~\ref{fig:theta_dist} shows the distributions of incident angles for both measured (un-weighted and weighted by the geometric correction factor) and initial photons. The distributions have a characteristic shape, peaking at $45^\circ$ and falling off at lower and higher angles, including a sharp cutoff at $90^\circ$. These characteristics can be well understood by considering the flux of a constant field through a hemisphere. Specifically, we consider the upper hemisphere from the setup shown in the left panel of Figure~\ref{fig:schematic}, with the positive $\hat{z}$ direction pointing towards the top of the page. The flux is calculated as
\begin{equation}
    \Phi = \iint_S \vec{F} \cdot \hat{n} \, dA,
\label{eq:flux_1}
\end{equation}
where $\vec{F}$ is the count rate from the surrounding sphere disk given by $\vec{F} = -F_0 \hat{z}$, with $F_0 = 0.074 \ \mathrm{ph \ km^{-2}}$. In spherical coordinates we have $\hat{n} = \hat{r}$ and $dA = R^2 \,  \mathrm{sin}\theta  \, d\theta \,  d\phi$, where $R$ is the radius of our watched volume, and $\theta,\phi$ are co-latitude and longitude, respectively. We can represent $\vec{F}$ in spherical coordinates as 
\begin{equation}
    \vec{F} = - F_0 \, (\mathrm{cos}\theta \, \hat{r} - \mathrm{sin} \theta \, \hat{\theta}).
\end{equation}
Plugging everything into Eq.~\ref{eq:flux_1}, and integrating over the azimuth angle, the magnitude of the flux through the hemisphere can now be written as
\begin{equation}
    \Phi = 2 \pi R^2 F_0 \int \mathrm{sin} \theta \, \mathrm{cos} \theta \,  d \theta,
\end{equation}
from which we obtain
\begin{equation}
\frac{d \Phi}{d \theta} (\theta) = 2 \pi R^2 F_0  \, \mathrm{sin} \theta \, \mathrm{cos} \theta \, d \theta.
\end{equation}
This equation is plotted with the dotted grey line in the left panel of Figure~\ref{fig:theta_dist}, and as can be seen, there is excellent agreement with the simulations. 

In order to more clearly exemplify the effect of the weights, in the right panel of Figure~\ref{fig:theta_dist} we show the distribution of measured incident angles, for an initial incident angle of $50^\circ$ and energy of 8 MeV, for both the weighted and un-weighted histograms. As can be seen, the weights are important for high off-axis angles, due to the effect of the projected area becoming increasingly smaller.

\section{Photon Interaction Sequences}
\label{sec:photon_interactions}

In this section we provide more insight into the different interactions that may occur during the photon transport through the atmosphere. As discussed in Section~\ref{sec:sims}, the possible interaction types include Compton scattering, pair conversion, photoelectric absorption, Bremsstrahlung radiation, and Rayleigh scattering. Correspondingly, secondary photons may also be produced from pair production and subsequent annihilation, as well as Bremsstrahlung radiation of electrons and positrons. In general, a photon may undergo any combination of interactions, and each of them can be tracked with our simulation pipeline\footnote{This comes at the cost of a larger simulation file, and so by default the pipeline does not track all the interaction information.}. An example of an interaction sequence for a given photon might be as follows: (1) The initial photon pair produces. (2) The generated electron emits a secondary photon via Bremsstrahlung radiation, which is then photo absorbed. (3) The generated positron annihilates producing two more secondary photons. (4) One of the secondary photons Compton scatters, and then enters the watched volume, becoming a ``detected" photon.

\begin{figure}
\centering
\includegraphics[width=0.32\columnwidth]{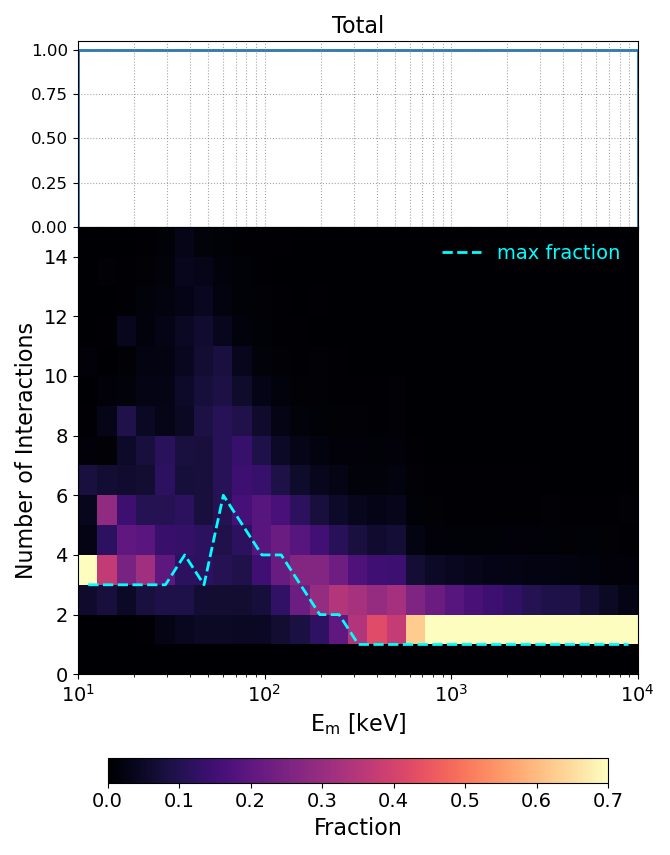}
\includegraphics[width=0.32\columnwidth]{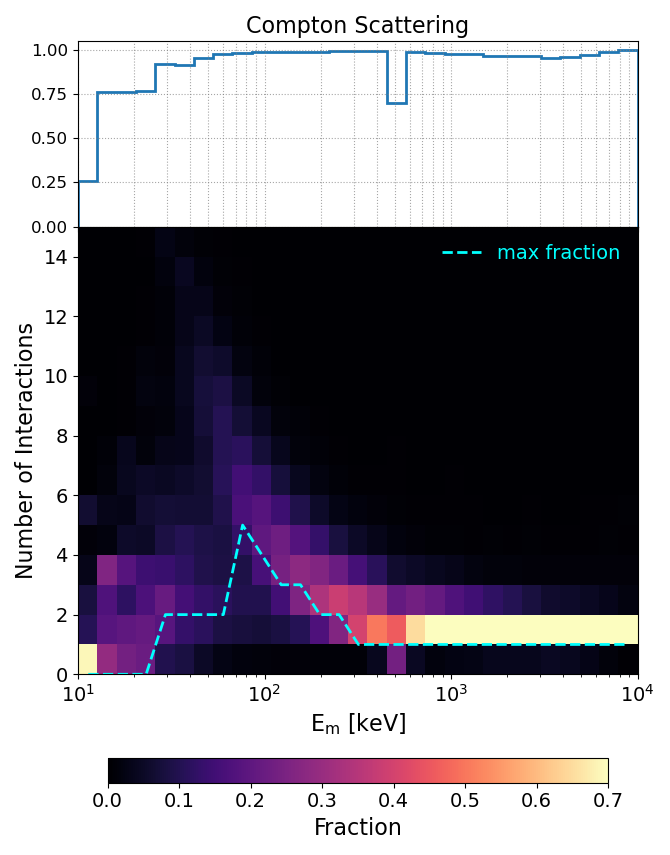}
\includegraphics[width=0.32\columnwidth]{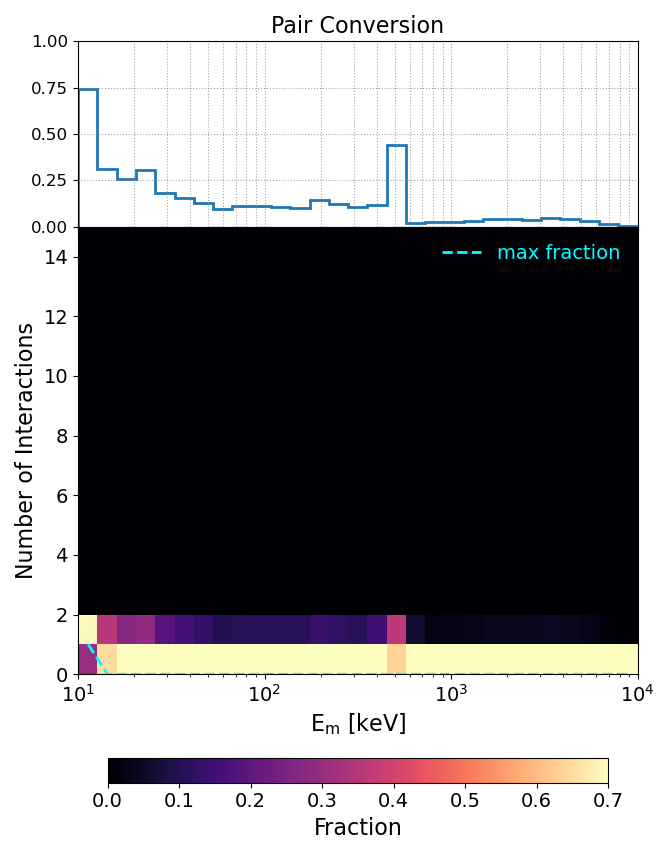}
\includegraphics[width=0.32\columnwidth]{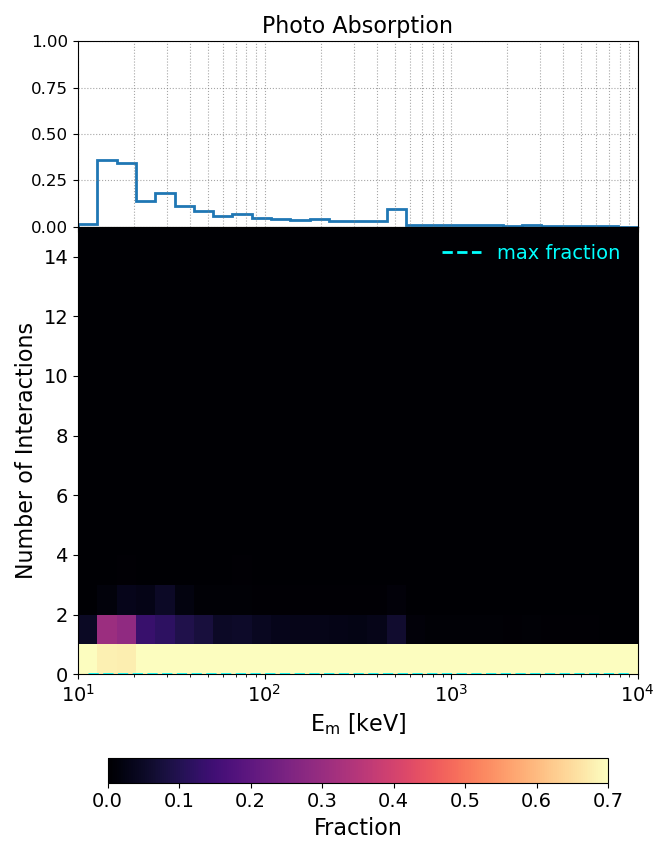}
\includegraphics[width=0.32\columnwidth]{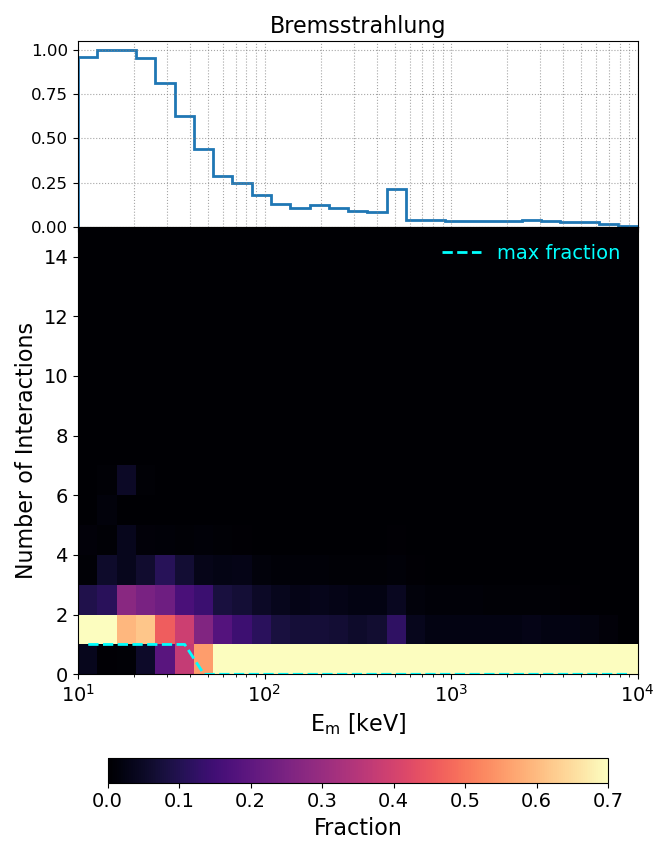}
\includegraphics[width=0.32\columnwidth]{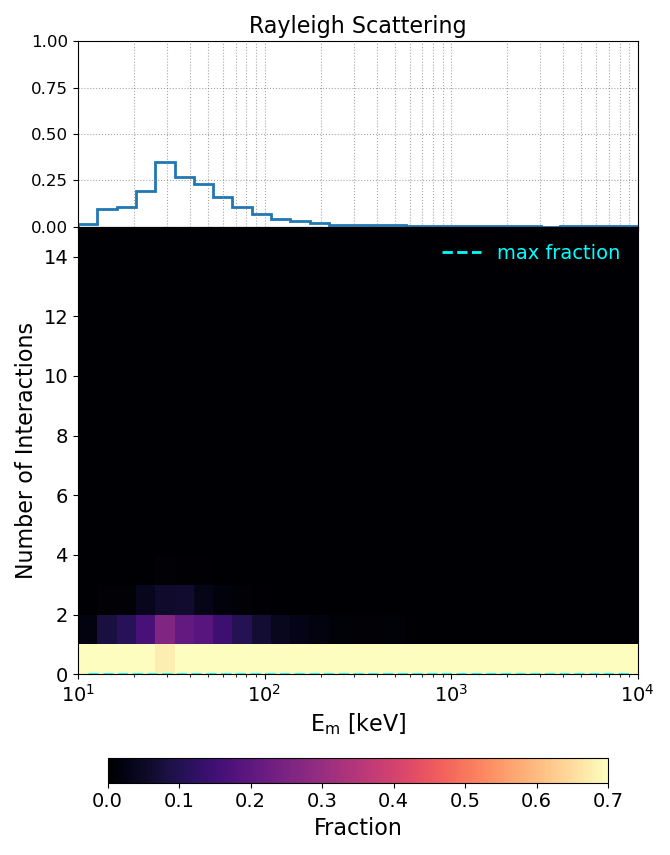}
\caption{The lower plot in each panel is a 2D histogram showing the number of interactions versus measured photon energy ($E_m$), where the interaction type is specified at the top of each respective panel. The histograms only contain photons that scatter at least once, and they are normalized by the total number of detected photons in each energy bin (which includes all interaction types). The colorbar is therefore normalized to 1, but the images are saturated at 0.7 in order to emphasize the distributions at lower energies. The dashed cyan lines show the maximum fraction as a function of energy. Directly above each 2D histogram is a 1D histogram showing the projection onto the measured energy axis. Note that when making the projection we do not include the zero bin for the number of interactions. Thus, the 1D histograms show the fraction of measured photons whose interaction sequence includes the respective interaction at least once, as a function of measured energy.}
\label{fig:interactions}
\end{figure}

In Figure~\ref{fig:interactions} we show the distributions of the actual interaction sequences from the simulations for all event types. Specifically, we show the number of interactions as a function of measured photon energy (for all incident angles). Note that these distributions include all the events that occurred in a photon interaction sequence before being measured, and so it does not necessarily mean that the measured photon was directly involved with each interaction (e.g., consider the sequence given above). We see that at higher energies (above $\sim$ 300 keV) a majority of the scattered photons undergo a single Compton interaction, with the exception of the 511 keV bin, which has a high fraction of pair events, as expected. Towards lower energies, the number of interactions that occur before the photon is measured increases, with the maximum fraction peaking at 6 interactions for a measured energy around $\sim$ 80 keV. This is due to an increased number of Compton scatters, as well as more interactions from pair conversion (and subsequent annihilation), photo absorption, Bremsstrahlung radiation, and Rayleigh scattering.

\section{Analytical Calculation of Transmission Probability}
\label{sec:tp_analytical}

\begin{figure*}[t]
\centering
\includegraphics[width=0.6\columnwidth]{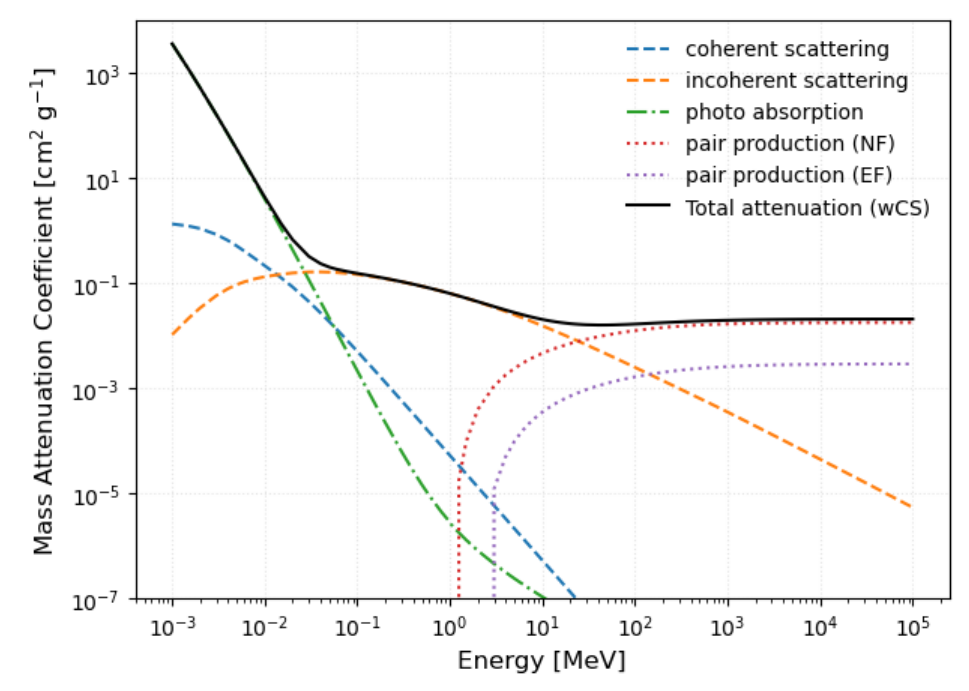}
\caption{Mass attenuation coefficients as a function of energy, based on an atmospheric composition consisting of 78\% $\mathrm{N}_2$ and 22\% $\mathrm{O}_2$. The different curves show different interactions, as specified in the legend. Pair production interactions are shown separately for interactions with a nuclear field (NF) and an electron field (EF). The total component is shown including coherent scattering (wCS).}
\label{fig:mass_att}
\end{figure*}

For a narrow beam of mono-energetic photons, the change in $\gamma$-ray beam intensity ($I$) at some distance ($x$) in a material can be expressed as:
\begin{equation}
   dI(x) = -I(x) \sigma n dx,
\end{equation}
where $\sigma$ is the interaction cross section (units of $\mathrm{cm}^2$), and $n$ is the number density of the material (units of $\#/\mathrm{cm}^3$). 
Integrating both sides along the line of sight gives:
\begin{equation}
    \int_{I_0}^{I_f} \frac{dI}{I} = \int_{x_0}^{x_f}-\sigma n\mathrm{d}x \
    \implies \frac{I_f}{I_0} = \mathrm{exp} \bigg( \int_{x_0}^{x_f}-\sigma n\mathrm{d}x \bigg). 
\end{equation}
We can replace the quantity $\sigma n$ with $\eta(E) \rho(x)$, where $\eta(E)$ is the mass attenuation coefficient as a function of energy (units of $\mathrm{cm^2/g}$), and $\rho(x)$ is the mass density (units of $\mathrm{g/cm^3}$). Thus, the TP can be calculated as
\begin{equation}
    \mathrm{TP} \equiv \frac{I_f}{I_0} = \mathrm{exp} \bigg( - \eta (E) \int_{x_0}^{x_f}\rho(x)dx \bigg).
\label{eq:tp}
\end{equation}
This calculation is implemented in the cosi-atmosphere package. The mass density is given by our atmospheric model. Note that the mass density is a function of radius, but this can easily be mapped to any point, $x$, along the line-of-sight, as given in Eq.~\ref{eq:tp} (e.g., using the law of cosines). Data for the mass attenuation coefficients are taken from the National Institute of Standards and Technology (NIST) XCOM\footnote{\url{https://physics.nist.gov/PhysRefData/Xcom/html/xcom1.html}} program. We use a simplified atmospheric mixture consisting of 78\% $\mathrm{N}_2$ and 22\% $\mathrm{O}_2$. This serves as a close approximation, particularly for altitudes $\lesssim $100 km. Figure~\ref{fig:mass_att} shows the mass attenuation coefficients as a function of energy, for different relevant interactions. As can be seen, incoherent scattering (i.e., Compton scattering) is dominant in our energy range of interest. Pair production dominates above $\sim30$ MeV, and photo absorption dominates below $\sim 50$ keV. Coherent scattering (i.e., Rayleigh scattering) also becomes important at low energies. 

\section{Spherical Mass Model: On-Axis Source}
\label{sec:on_axis_spherical}
\begin{figure*}
\centering
\includegraphics[width=1\columnwidth]{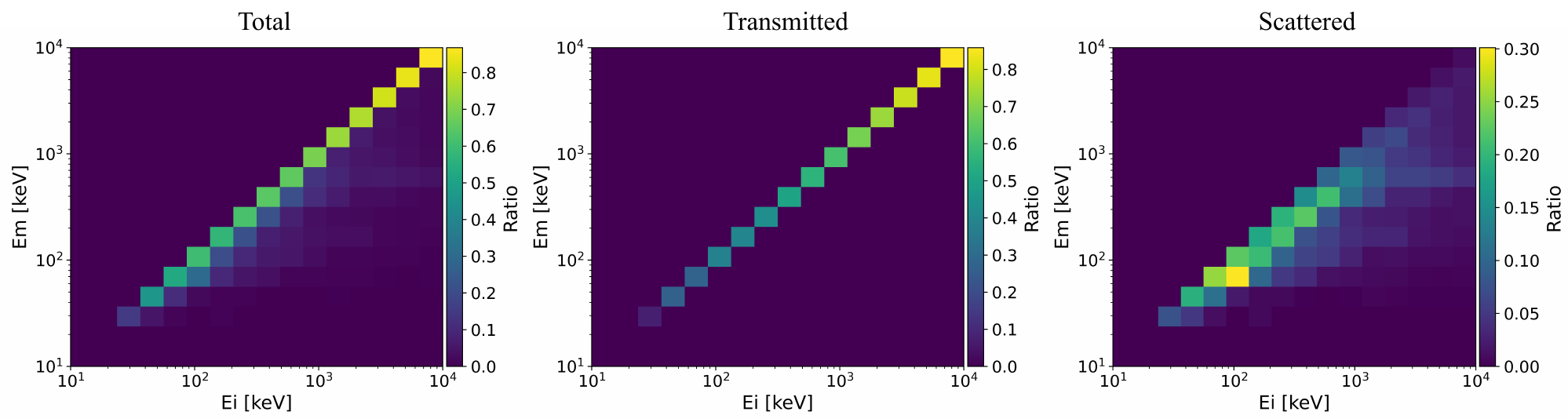}
\caption{Energy dispersion matrix for all photons (left), transmitted photons (middle), and scattered photons (right), for an on-axis source. The y-axis is the measured photon energy, and the x-axis is the initial photon energy. The matrices are normalized by the total photons simulated in each bin of $E_i$ and $\theta_i$. The total energy dispersion matrix is the sum of the transmitted and scattered components.}
\label{fig:energy_dispersion_on_axis_sphere}
\end{figure*}

\begin{figure*}
\centering
\includegraphics[width=0.32\columnwidth]{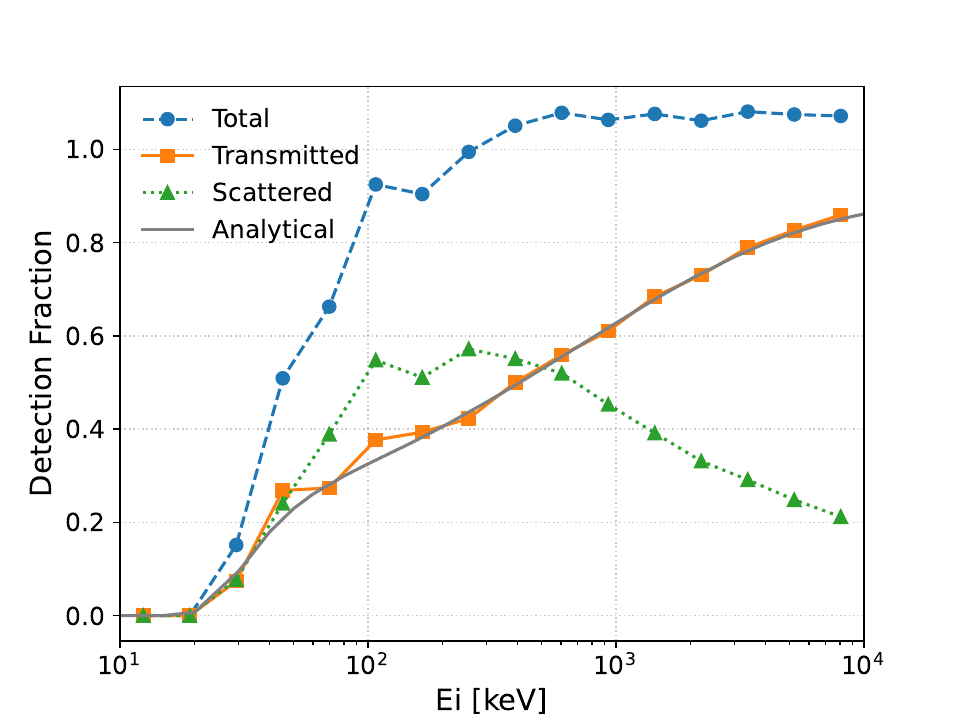}
\includegraphics[width=0.32\columnwidth]{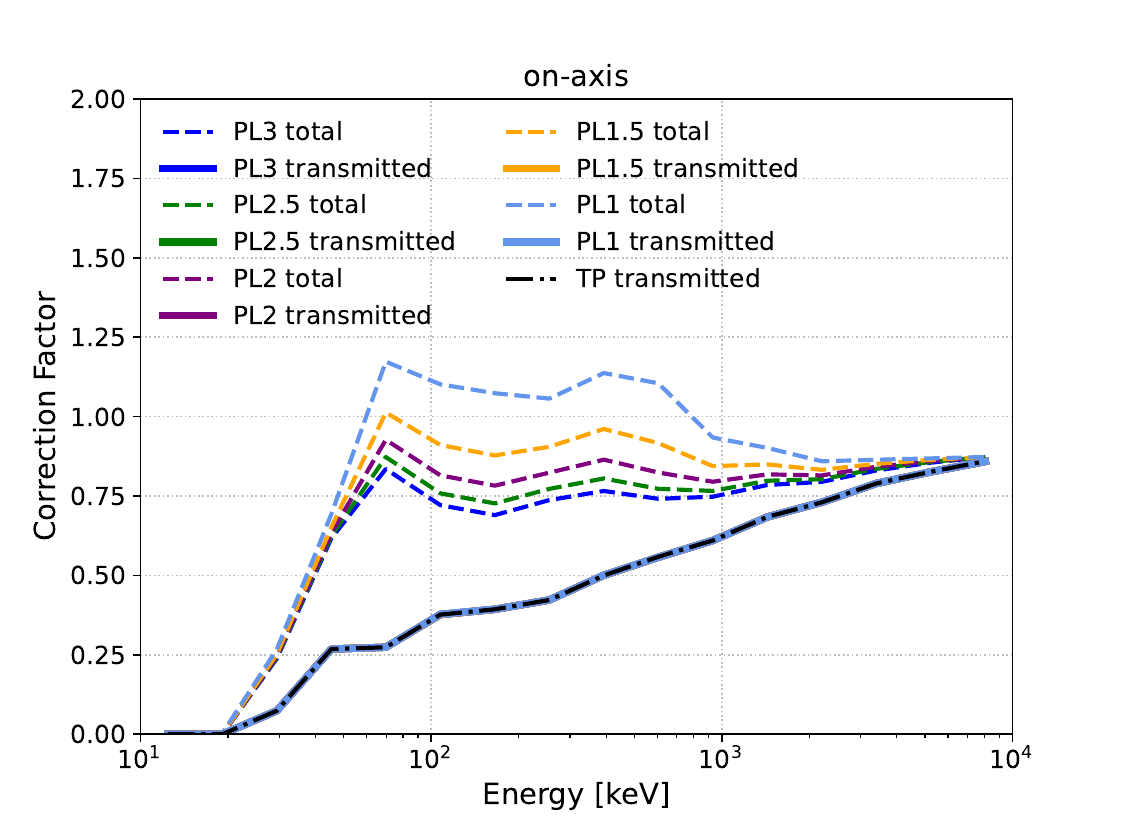}
\includegraphics[width=0.32\columnwidth]{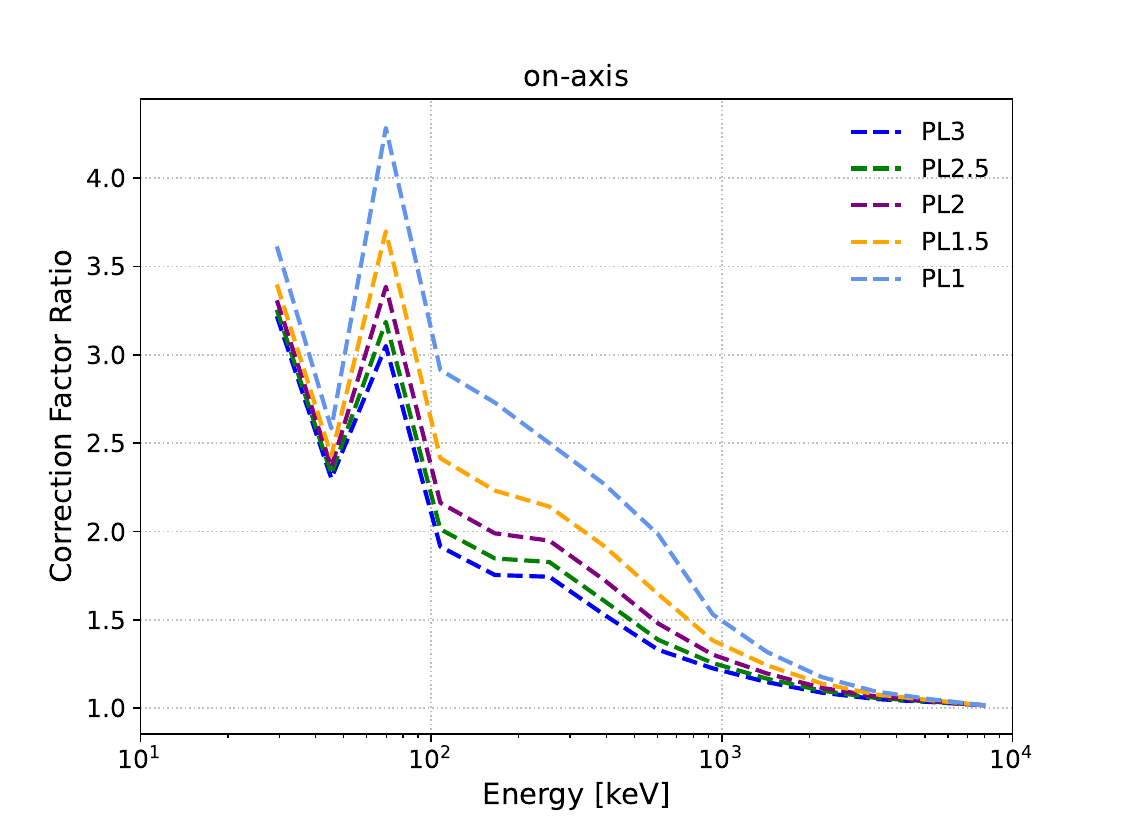}
\caption{\textbf{Left:} Transmission probability for the three components, as specified in the legend. For reference, the TP calculated analytically is also overlaid. \textbf{Middle:} Correction factor for a range of power law spectral models ranging form 1 $-$ 3. The solid line is for the transmitted component only, and the dashed line is for the total component, which includes transmitted and scattered photons. The black dash-dot line is the TP, obtained by projecting the energy dispersion matrix onto the initial energy axis. \textbf{Right:} Correction factor ratio for the same range of models shown in the left plot.}
\label{fig:TP_on_axis_sphere}
\end{figure*}

Figure~\ref{fig:energy_dispersion_on_axis_sphere} shows the energy dispersion matrices for an on-axis source. Qualitatively, they are consistent with the results for the $50^\circ$ off-axis source presented in Section~\ref{sec:edisp}. The corresponding detection fraction is shown in the left panel of Figure~\ref{fig:TP_on_axis_sphere}. Again, they are qualitatively consistent with the $50^\circ$ off-axis case. The main difference is that the detection fraction is overall higher for the on-axis source, as expected. We note that some statistical variation is evident in the correction factor curves. Indeed, the on-axis case has the lowest statistics, as shown in Figure~\ref{fig:theta_dist}. Further simulations should resolve this and produce smooth curves. In the middle and right panels of Figure~\ref{fig:TP_on_axis_sphere} we show the corresponding correction factor and correction factor ratio, respectively.

\section{Rectangular Mass Model}
\label{sec:rectangular_mass_model}

As an alternative approach to the spherical mass model of the atmosphere, a rectangular geometry can be employed. In this setup, instead of using spherical atmospheric shells, the atmosphere is described using planar slabs. Additionally, a narrow beam is used instead of an isotropic source. Otherwise, the atmosphere is still modeled using NRLMSIS, and the response is defined and analyzed in a similar way.

\begin{figure*}
\centering
\includegraphics[width=0.55\columnwidth]{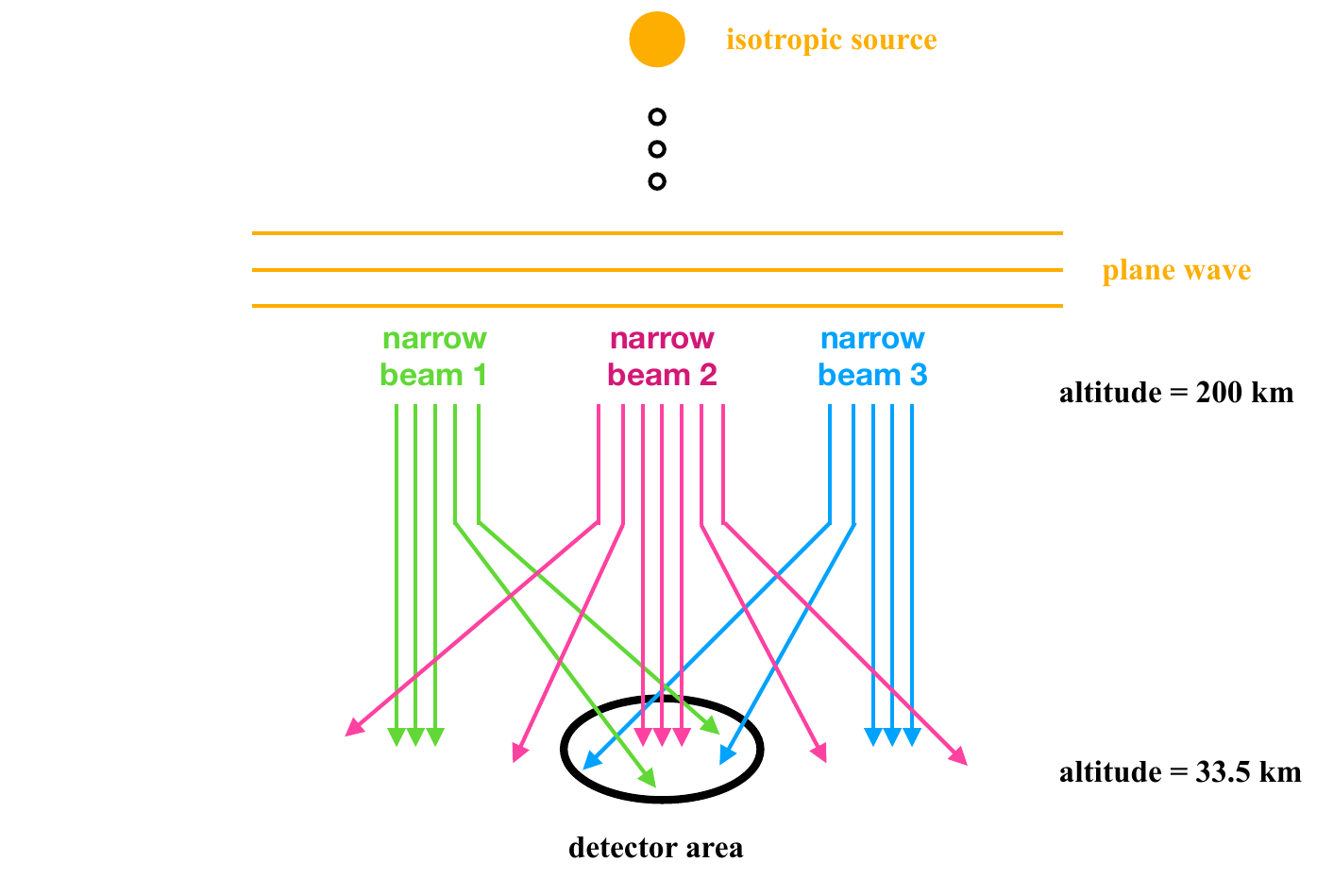}
\caption{Schematic depicting the setup of the rectangular mass model, as described in the text.}
\label{fig:rectangle_schematic}
\end{figure*}

\begin{figure}
\centering
\includegraphics[width=0.32\columnwidth]{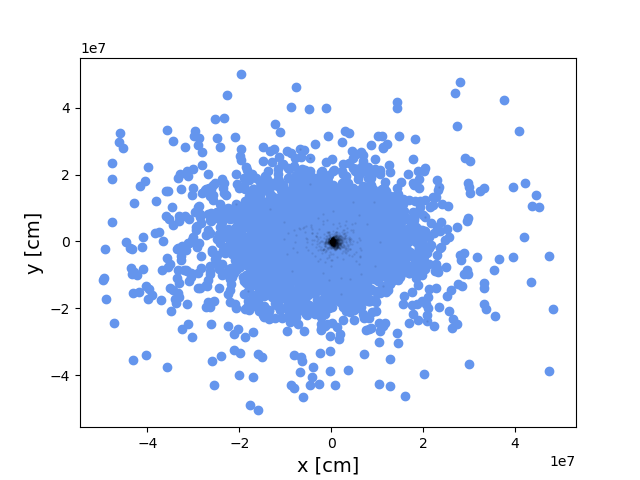}
\includegraphics[width=0.32\columnwidth]{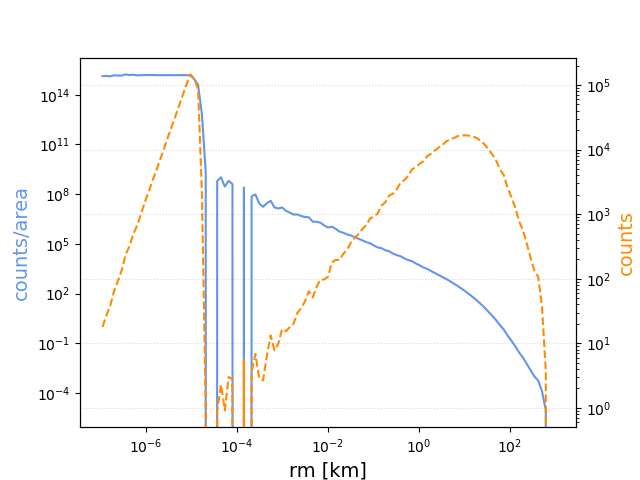}
\includegraphics[width=0.32\columnwidth]{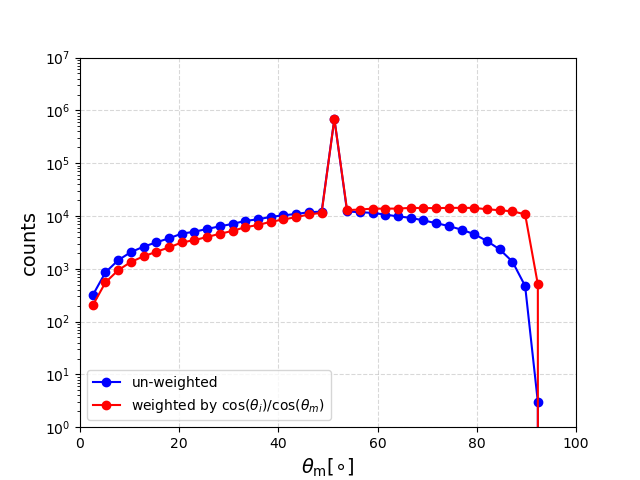}
\caption{\textbf{Left:} 2-dimensional scatter plot of measured photon positions, for a $50^\circ$ off-axis source, using a rectangular mass model. The black circles show the measured positions of the transmitted photons, and the blue circles show the measured positions of all photons (transmitted + scattered). \textbf{Middle:} Radial distribution of the measured photons. The left y-axis shows counts/area (corresponding to the blue curves), and the right y-axes shows the total counts (corresponding to the orange curve). \textbf{Right:} Angular distribution of the measured photons. For reference, we show both weighted and un-weighted histograms.}
\label{fig:rect_50deg_dist}
\end{figure}

The logic behind the rectangular simulations can be understood by considering the schematic in Figure~\ref{fig:rectangle_schematic}. A typical (non-beamed) astrophysical source radiates isotropically. The curvature of the wave decreases with the square of the distance from the source, and thus it can be approximated as a plane wave once it eventually passes Earth. The plane wave can be described as the superposition of many narrow beams. Three such beams are depicted in Figure~\ref{fig:rectangle_schematic}. If we consider the middle beam (narrow beam 2), some photons from the beam will reach the detector without scattering, and other photons will scatter and never reach the detecting area. This corresponds to the transmitted component, as we have defined in this work. If we consider the two adjacent beams (narrow beams 1 and 3), the same thing also occurs. Thus, the same distribution of photons that scatter out of the detector area also scatter back into it, from the superposition of all nearby beams. This is simply a consequence of the circular geometric symmetry of the problem. The scattered photons that enter the detector correspond to the scattered component, as we have also defined in this work. By watching the entire plane at 33.5 km, we can characterize the scattered photons using a single narrow beam source (radius of 1 cm) placed at 200 km.

We simulate a $50^\circ$ off-axis source using the rectangular mass model. Photons are classified as having scattered if their measured incident angle varies by more than $0.2^\circ$ from the initial incident angle. Note that the angular resolution here is much better than that used for binning the spherical mass model. The left panel of Figure~\ref{fig:rect_50deg_dist} shows the positions of the measured photons, where we show separately the transmitted and total (transmitted + scattered) components. Correspondingly, the middle panel of Figure~\ref{fig:rect_50deg_dist} shows the radial distribution of the measured photons, and the right panel shows the distribution of measured incident angles. The radial distribution shows a bimodal feature. First, there is a flat distribution (in counts/area) between $\sim 0-1$ cm, corresponding to the beam radius of the simulated photons. This part of the distribution corresponds to the transmitted component. The second part of the distribution corresponds to the scattered component, and the intensity gradually falls off with increasing radius. In terms of total counts, the scattered component peaks near $\sim 10$ km, before quickly falling off. This indicates that a majority of the photons the contribute to the scattered component originate from within $\sim 10$ km of the detector. Because the radius of Earth is so large, the curvature of its surface within 10 km is minimal\footnote{Moving along a flat surface compared to a spherical surface within a 10 km radius leads to an altitude difference ($\Delta h$) of $\Delta h / R_E = 10^{-6}$.}, and thus we can approximate the surface as being locally flat in the region where a majority of the scattered photons originate from. For this reason, we expect the rectangular mass model to give comparable results as those obtained with the spherical mass model, and indeed, this is what we find. 

\begin{figure*}
\centering
\includegraphics[width=1\columnwidth]{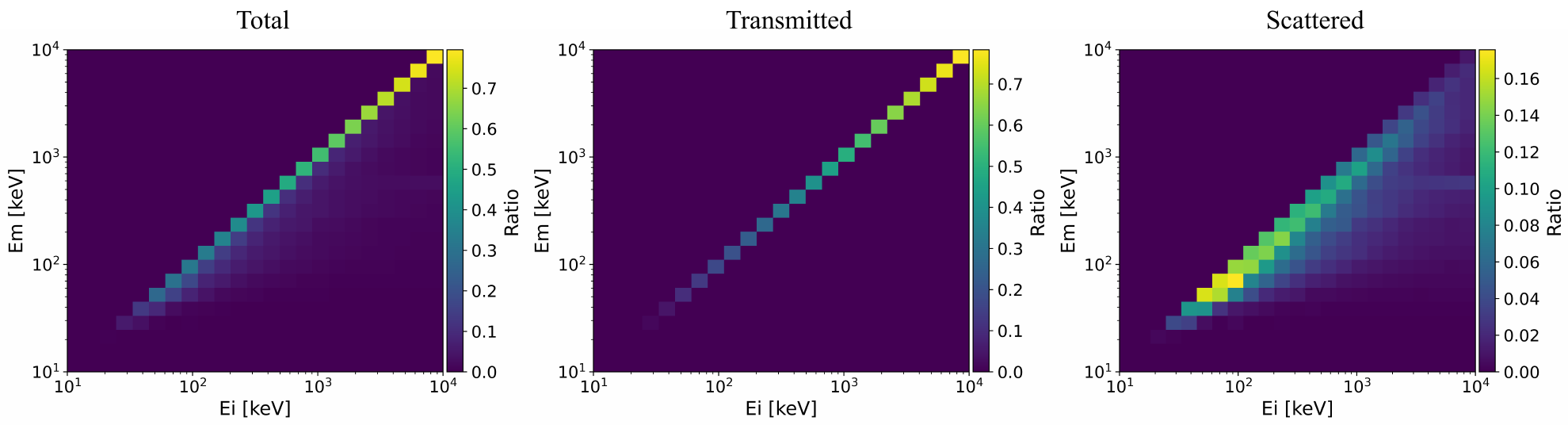}
\caption{Energy dispersion matrix for all photons (left), transmitted photons (middle), and scattered photons (left), for a 50$^\circ$ off-axis source, using a rectangular mass model. The y-axis is the measured photon energy (Em), and the x-axis is the initial photon energy. The matrices are normalized by the total photons simulated in each energy bin. The total energy dispersion matrix is the sum of the transmitted and scattered matrices.}
\label{fig:energy_dispersion_50deg_rect}
\end{figure*}

\begin{figure}
\centering
\includegraphics[width=0.32\columnwidth]{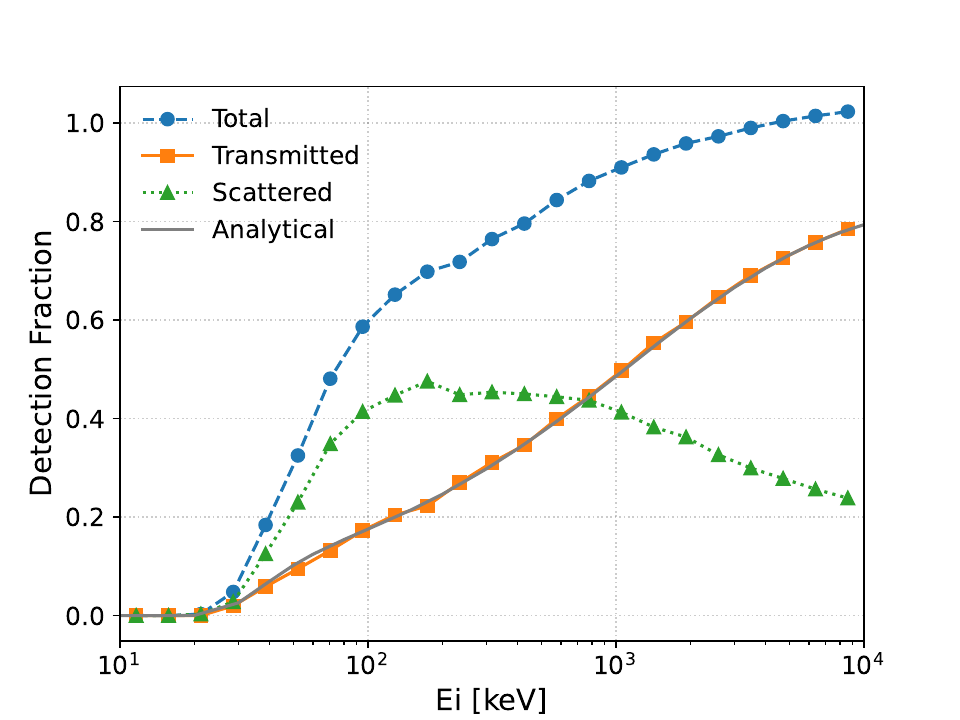}
\includegraphics[width=0.32\columnwidth]{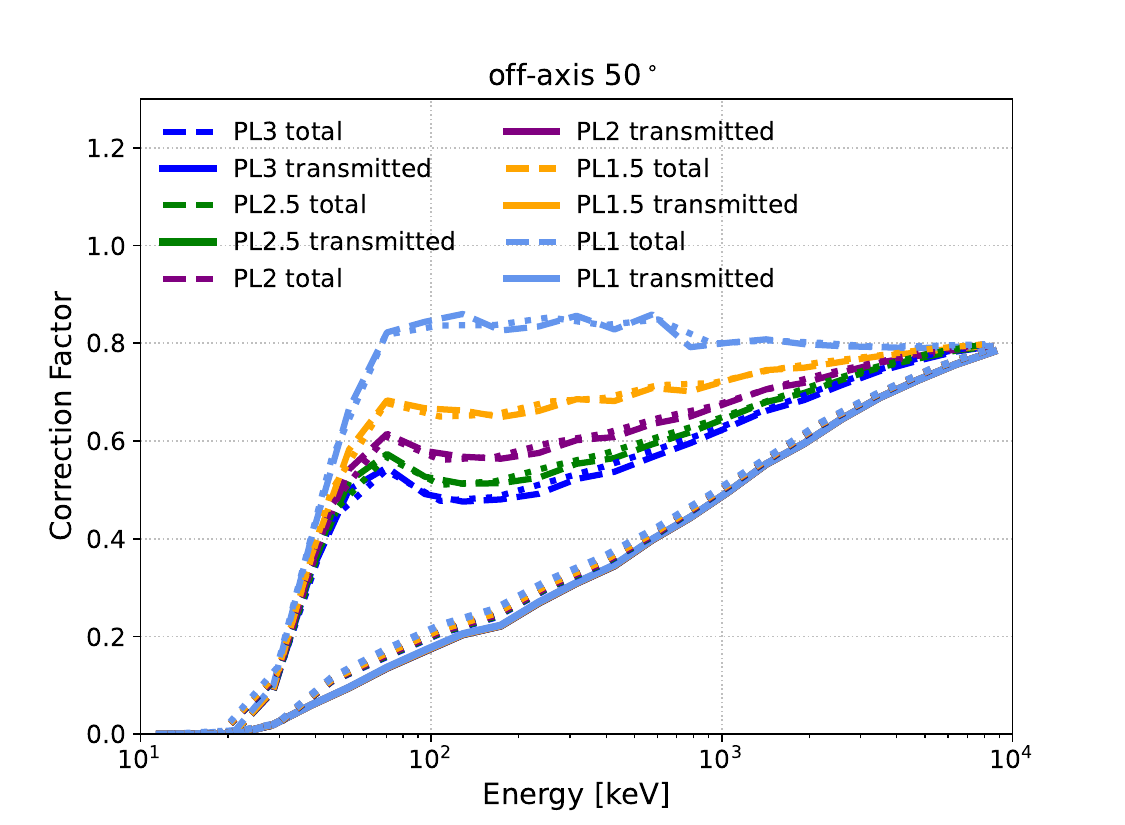}
\includegraphics[width=0.32\columnwidth]{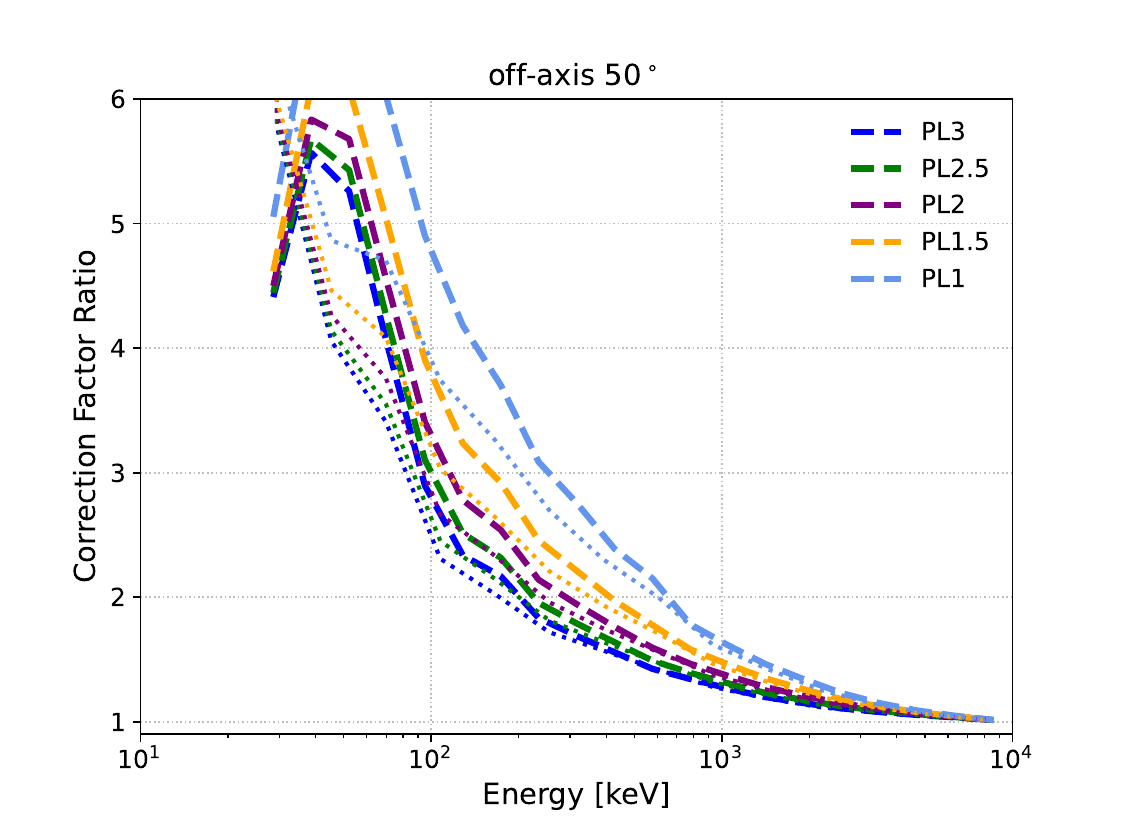}
\caption{\textbf{Left}: Detection fraction for the three components, as specified in the legend, for a 50$^\circ$ off-axis source, using a rectangular mass model. For reference, the TP calculated analytically is also overlaid. \textbf{Middle:} Correction factor for power law spectral models ranging form $1-3$, for a 50$^\circ$ off-axis source, using a rectangular mass model. The solid line is for the transmitted component only, and the dashed line is for the total component, which includes transmitted and scattered photons. The dotted curves show the results from the spherical mass model. \textbf{Right:} Correction factor ratio for the same range of models shown in the left plot.}
\label{fig:tp_on_axis}
\end{figure}

Figure~\ref{fig:energy_dispersion_50deg_rect} shows the energy dispersion matrices. They are very similar to the results with the spherical mass model. Note that with this simulation setup it is easier to obtain high statistics because we simulate a single initial incident angle at a time, which allows for finer resolution. The detection fraction is shown in the left panel of Figure~\ref{fig:tp_on_axis}. The middle and right panels of Figure~\ref{fig:tp_on_axis} show the correction factor and correction factor ratio, respectively. Overlaid in these plots are the corresponding results from the spherical mass model. As can be seen, they are generally in very good agreement. The total correction factors match well, although the transmitted component is slightly higher at low energy for the spherical mass model. This is most likely due to the coarser binning (in angle and energy) that is used for the spherical mass model. Likewise, the correction factor ratio for the rectangular mass model is higher towards low energy, which can be attributed to the difference in the binning. 

\begin{figure}
\centering
\includegraphics[width=0.32\columnwidth]{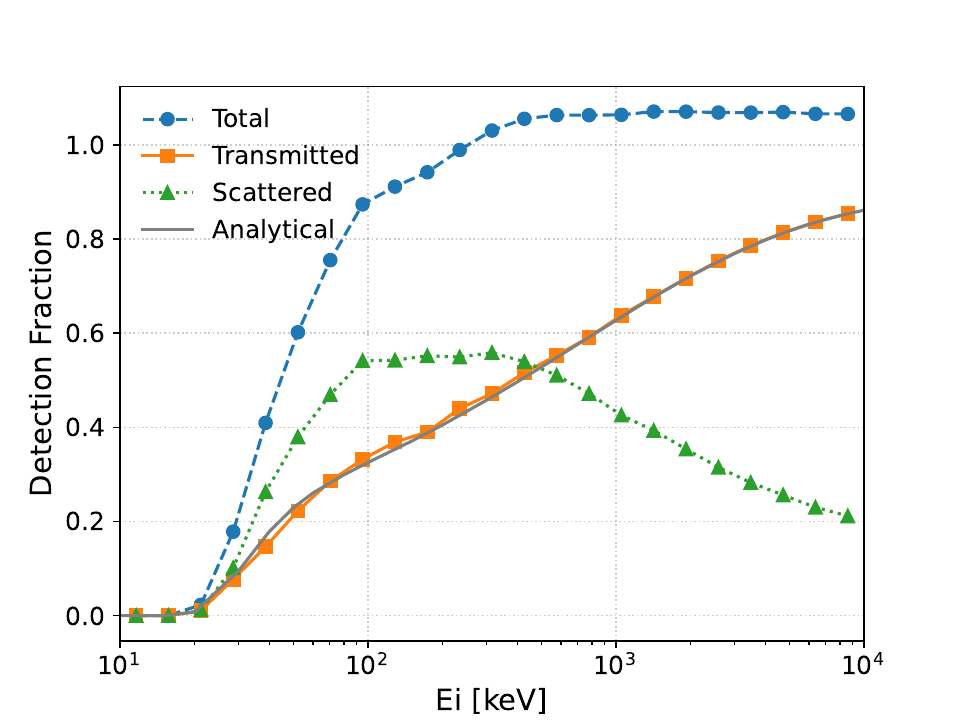}
\includegraphics[width=0.32\columnwidth]{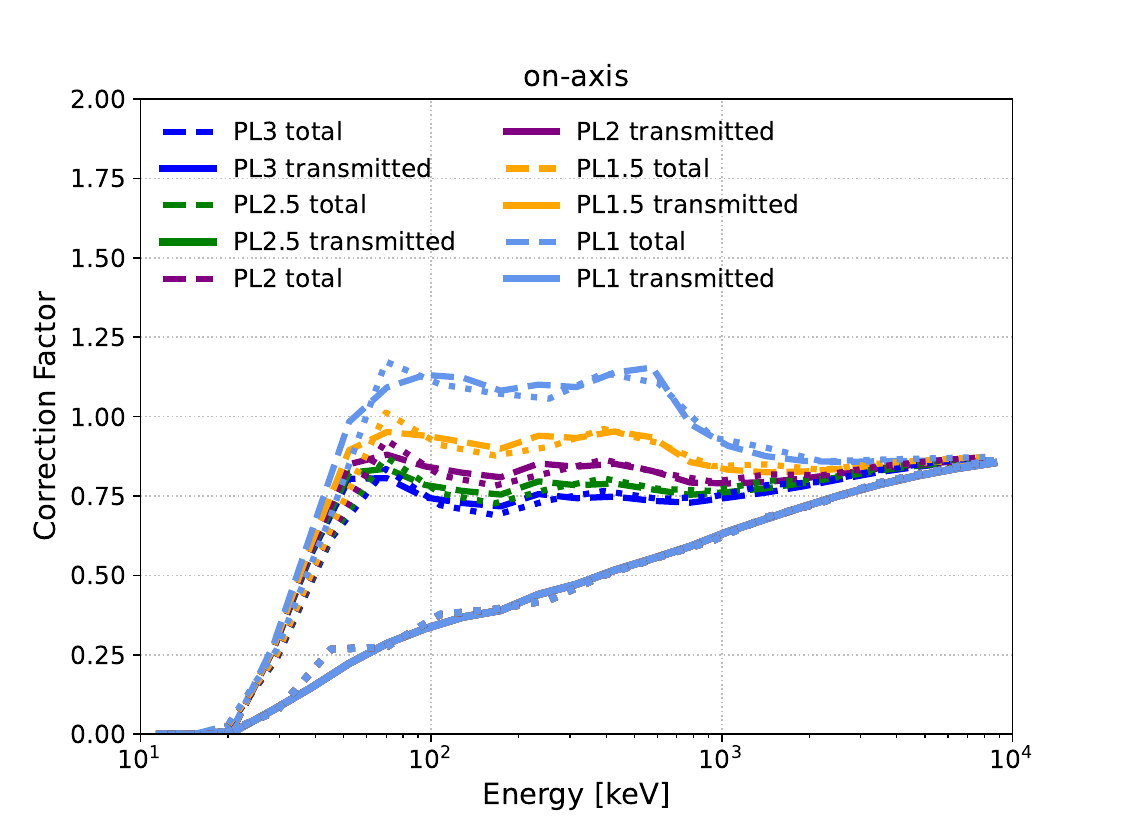}
\includegraphics[width=0.32\columnwidth]{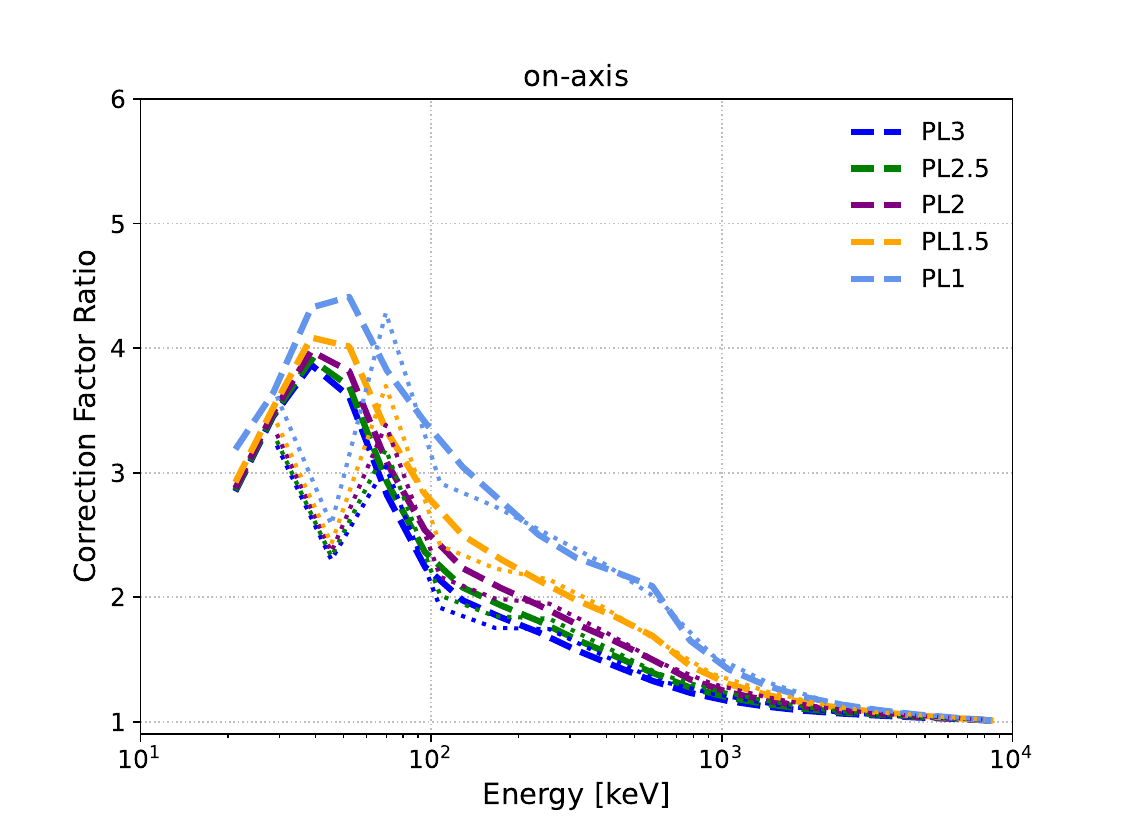}
\caption{\textbf{Left}: Detection fraction for the three components, as specified in the legend, for an on-axis source, using a rectangular mass model. For reference, the TP calculated analytically is also overlaid. \textbf{Middle:} Correction factor for power law spectral models ranging form $1-3$, for an on-axis source, using a rectangular mass model. The solid line is for the transmitted component only, and the dashed line is for the total component, which includes transmitted and scattered photons. The dotted curves show the results from the spherical mass model. \textbf{Right:} Correction factor ratio for the same range of models shown in the left plot.}
\label{fig:rect_on_axis}
\end{figure}

As a further example, Figure~\ref{fig:rect_on_axis} shows the correction factors and correction factor ratios for an on-axis source. Again, there is very good agreement between the rectangular and spherical mass models. This overall agreement serves as another validation of the spherical geometry simulations, and at the same time, it also shows that the rectangular geometry can be used as an alternative/complementary approach to approximating the atmospheric response. For example, this would be particularly useful to study the behaviour of resolved $\gamma$-ray lines, for which more statistics in much smaller energy bins is required.

\pagebreak
\bibliography{sample631}
\bibliographystyle{aasjournal}



\end{document}